\documentstyle[floats,aps,epsfig,eqsecnum,amsfonts,amssymb]{revtex}
\newcommand{\mathbold}[1]{\mbox{\boldmath $#1$}}

\title{Fluctuations of elastic interfaces in 
      fluids: Theory and simulation}

\author{Davide Stelitano}

\address{Department  of Physics\\
         Massachusetts Institute of Technology, 
         Cambridge, MA 02139}

\author{Daniel H. Rothman}

\address{Department of Earth, Atmospheric and Planetary Sciences\\
         Massachusetts Institute of Technology,
         Cambridge, MA 02139}

\date{\today}
\begin{document}

\wideabs{
\maketitle
\begin{abstract}
We study 
the  dynamics of elastic interfaces---membranes---immersed
in thermally excited fluids. The work contains three
components: the development of a numerical method,
a purely theoretical approach,
and numerical simulation.
In developing a numerical method, we first 
discuss the dynamical coupling between the 
interface and the surrounding fluids. An argument is then 
presented that generalizes the single-relaxation time
lattice-Boltzmann method for the simulation of hydrodynamic
interfaces to include the elastic properties of the 
boundary.
The implementation of the new method is 
outlined and it is tested by simulating the
static behavior of spherical bubbles and 
the dynamics of bending waves.
By means of the fluctuation-dissipation theorem we recover 
analytically the equilibrium frequency power spectrum 
of thermally fluctuating membranes and the correlation
function of the excitations. Also, the non-equilibrium
scaling properties of the membrane roughening are
deduced, leading us to formulate a scaling law describing
the interface growth, $W^2(L,T)=L^3\,g(t/L^{5/2})$,
where $W$, $L$ and $T$ are the width
of the interface, 
the linear size of the system and the temperature 
respectively, and $g$ is a scaling function.
Finally, the phenomenology of thermally fluctuating membranes
is simulated and the frequency power spectrum is 
recovered, confirming the decay of the correlation function of
the fluctuations. As a further 
numerical study of fluctuating elastic interfaces,
the non-equilibrium
regime is reproduced by initializing the system as an interface
immersed in thermally pre-excited fluids.
\end{abstract}
}

\section{Introduction}

Elastic interfaces in fluids, such as biological membranes, have 
spurred a strong interest  in recent years not only because of their 
practical importance, but also because of the intriguing 
complexity of their phenomenology.
Many aspects of this rich subject have been studied
from different approaches. Since  temperature and
its effects play a primary role in natural phenomena,
thermal fluctuations of fluid, hexatic, nematic and 
polymerized membranes have been theoretically discussed
in a number of works
\cite{land1,broc,arono,pow1,gomp1,dia}.
Also, due to their relevance to mechanical and chemical
interactions in biological systems, 
shape transformations and fluctuating topologies
of elastic interfaces immersed in fluids with, in many cases, 
shear flows have been thoroughly studied 
\cite{rama,goz1,navo,eggle,gomp2,goz2}.
Crystalline membranes, also known as polymerized or tethered 
membranes, are, in particular, a fascinating subject 
with important concrete
realizations in nature, such as the cytoskeleton of mammalian
erythrocytes (red blood cells)~\cite{elgsa,schmi}.
Among the systems that can be physically realized in a laboratory,
inorganic crystalline membranes were 
examined in~\cite{tan1,tan2}. 
On the other hand, theoretical work on
tethered membranes has proved 
successful in
addressing the crumpling 
transition~\cite{kard1,kard2,kard3,david,guitt}.
Other theoretical results on finite-size effects in fluid membranes 
can be found in~\cite{miln1,miln2,morse1}.
Furthermore, some books and reports 
\cite{safra,lipo,nels,david2,bowi}
reviewing the statistical
mechanics, thermodynamics and geometrical structure
of membranes have
shed further light on the understanding of the subject.

In the present work, we focus on the physics of
elastic interfaces in fluids when thermal 
fluctuations in the bulk generate correlated forces and subsequent
excitations on the boundary. This dynamical coupling between
interface and surrounding fluid is believed to be responsible
for interesting phenomena, such as the flickering of
erythrocytes~\cite{broc} or the physical distribution
of particulates inside certain lipid vescicles~\cite{tony}.
Because many such problems do not have a closed-form solution,
numerical simulations can provide us with a valuable tool for
a deeper understanding and general guidelines for 
designing new experiments.
More generally, the ability to simulate thermally fluctuating 
elastic membranes in fluids allows a computational 
model to be effectively employed in the simulation of the biophysical
systems for which the effects of a finite temperature
are to be accounted for.
Computational works on fluctuating membranes
have appeared rarely in literature 
\cite{gomp2,gomp3,gonn,gomp4}. 
As an interesting example, 
in the work by Goetz {\it et al.}~\cite{gomp3}
the power spectrum of 
a fluctuating bilayer membrane in vacuum
is obtained by
molecular dynamics. 
Also, the reader may find in~\cite{gonn} an example of how a 
mean-field theory approach in the framework of a 
lattice-Boltzmann model can be effectively employed in
the study of phase separation with boundaries
driven by surface tension and bending stiffness.
Broadly
speaking, elastic forces are governed by the local curvature
of the interface and their wavelengths are two to four orders of
magnitude larger than molecular ones. 
Therefore, a major difficulty
in modeling such systems is one of describing 
the problem at the different length and time scales 
of molecular and elastic interactions in an unified and 
consistent approach.

Our work is threefold. First, we create a new
numerical method for the study of fluctuating membranes.
Second, we use the method to simulate phenomena 
associated with the coupling of the interface and the surrounding 
fluids. Third, after reviewing the properties of
fluctuating membranes at equilibrium, we deduce the 
non-equilibrium scaling law  
governing the interface roughening
and show that our predictions are observed
in simulations.

\par We develop a lattice-Boltzmann model.
Not only does our 
method allow the simulation of bending waves and 
fluctuating membranes in fluids, but it also allows the study of 
more complicated physical 
problems, in which the interface has many distinct components and the 
fluids have prescribed velocity vector fields, inducing 
stresses on the boundary.
So that our method may eventually be used to 
simulate complex flows in complex geometries, 
it is designed to produce thin interfaces, 
with a thickness of the order of
few ($\sim3$) lattice units. 
Accordingly, we choose to not describe the 
elastic boundaries by means of 
a slowly varying order parameter. 
We choose instead to represent the dynamics of interfaces 
with bending rigidity by means of a free energy
\cite{land1,gomp2,helf}, in which the location and 
the geometrical properties---the curvature---of 
the membrane appear explicitly.
Thermal fluctuations are introduced in the model as a Gaussian 
noise 
in the lattice-Boltzmann equation~\cite{ladd1,ladd2}. 
The link between differential 
geometry and microscopic dynamics is provided by a suitable 
perturbation, driven by the local curvature, of the occupation 
numbers, similarly to a procedure already successful in the
study of interfaces with surface tension~\cite{roth1}.

This paper is organized as follows.
In Section \ref{interdyn} we present an overview
of the relevant interface and fluid dynamics.
Section \ref{sub:micdyn}
focuses on the fluid-interface coupling and on how 
the macroscopic equations of motion for the membrane
are translated into microscopic mechanical prescriptions
for the occupation numbers. 
We thus provide the 
theoretical basis for building a lattice-Boltzmann
computational model.
Section \ref{test} reports the results of the 
simulation of spherical bubbles and bending waves.
Also, the experimental dispersion relation is here 
compared to the theoretical prediction.
In the following section we then discuss
the theoretical description of the physics
of one-dimensional fluctuating membranes
coupled to thermally excited  fluids.
In Section \ref{thermy} we use our model to simulate 
fluctuating elastic interfaces and study 
both the non-equilibrium roughening and 
the stationary state. We present here the results of our
computations and compare them with the theory
previously outlined. Conclusions 
follow in Section \ref{conc}.

\section{Dynamical coupling of elastic interfaces
         to an external force}\label{interdyn}

In this section we provide a theoretical motivation
for the lattice-Boltzmann microscopic dynamics 
that constitutes the basis of our model.
We recall the common expression of the 
free energy for elastic interfaces and apply
Hamilton's variational principle to recover the 
macroscopic equation of motion.

The dynamics of membranes with bending or flexural rigidity 
$\epsilon$
and surface tension $\sigma$ is assumed to be governed, 
for the case of a vanishing spontaneous curvature, 
by the free energy~\cite{land1,gomp2,helf}
\begin{equation}
\label{freen}
{\cal{F}}={1 \over 2} \epsilon\int{dS H^2} + 
\bar{\epsilon}\int{dS K} + \sigma\int{dS},
\end{equation}
where the integrations are performed over the area of a
2D membrane or over the length of a 1D interface. 
$H$ is the sum of the principal curvatures and $K$ is their 
product (the Gaussian curvature).
$\bar{\epsilon}$ is named the saddle-splay modulus and
the second term
in the above equation is responsible for the energy gain/loss 
due to a change in the interface genus, according to the 
Gau\ss-Bonnet
theorem 
$\int{dS K}=2\pi\chi=4\pi(1-g)$, where $\chi$ is the 
Euler-Poincar\'{e} characteristic of the surface and $g$ is
its genus~\cite{cher}.
For the 1D interfaces  studied here, this term acts as a spontaneous
curvature. Since we set it to vanish, the saddle-splay
term will be neglected.
The dynamical coupling of the membrane with its surroundings is
realized by introducing the force ${\mathbold F}$ per unit 
area or unit length exerted by the fluids on the interface.
By applying Hamilton's variational principle
to the free energy (\ref{freen})
and including a term for the work done by ${\mathbold F}$ when the 
membrane undergoes a configurational change~\cite{land1,edwa,harr}, 
the  following relation 
between $F_{\perp}$, the component of the
force perpendicular to the interface, and the
geometric properties is derived in Appendix \ref{appendixa}
\begin{equation}
F_{\perp}= \sigma H + \epsilon H \sum{\gamma_i^2} - 
\epsilon \nabla^2 H \,.
\end{equation}
Here the $\gamma_i$'s are the principal curvatures of the 
($n-1$)-dimensional
hypersurface.
This expression specializes to the case of a 1D membrane
as
\begin{equation}
\label{1dim}
F_{\perp}= \sigma \gamma + \epsilon \gamma^3  - 
\epsilon \frac{d^2\gamma}{ds^2}. 
\end{equation}
In (\ref{1dim}) $s$ is the arc length of a canonical 
parameterization of the interface and $\gamma(s)$ is the
local curvature.

\section{Numerical method}\label{sub:micdyn}

The lattice-Boltzmann method~\cite{roth1,doole} 
we adopt solves the incompressible
Navier-Stokes equations for the fluid dynamics, namely
\begin{equation}
\label{eqn: navisto}
\begin{array}{cl}
\rho \left( \partial_t {\bf u} + {\bf u}\cdot 
{\mathbold \nabla}{\bf u}\right)=\mu\nabla^2{\bf u}-
{\mathbold \nabla}p\\
{\mathbold \nabla}\cdot{\bf u}=0.
\end{array}
\end{equation}
Here ${\bf u}$ is the fluid macroscopic velocity,
$p$ is the local pressure, $\rho$ the density and $\mu$
is the viscosity coefficient.
We show in this section how 
${\bf u}$, $p$ and $\rho$ are expressed in terms of 
microscopic quantities and how the coupling 
between the fluid dynamics (\ref{eqn: navisto})
and the interface dynamics (\ref{1dim}) is realized 
by means of the microscopic pressure tensor.

\par In making the connection between the macroscopic relation 
(\ref{1dim})
and the microscopic dynamics of our lattice-Boltzmann 
method, we 
shall follow a procedure that has been previously 
employed 
for interfaces
with surface tension
\cite{roth1,roth2}.
Notice first that $F_{\perp}$ corresponds to the local 
fluid pressure gap 
$\Delta p =p_1 - p_2$ across the interface, so that one 
can rewrite 
(\ref{1dim}) as
\begin{equation}
\label{deltap}
\Delta p(s)= \sigma \gamma(s) + \epsilon \gamma^3(s)  - 
\epsilon \ddot{\gamma}(s),
\end{equation}
where a dot denotes the derivative with respect to the arclength.
The mechanical relation~\cite{roth1,rowl} between the normal pressure 
and the 
longitudinal pressure $p_t$ reads
\begin{equation}
\label{mechrel}
\Delta p(s)= \gamma(s)\int_{-\infty}^{\infty}
{\left[\bar{p}_n(s)-p_t(s,y)\right]dy}
\end{equation}
where the integration is carried out along the $y$ direction 
perpendicular to the interface.
The average between the pressures at either side of the 
membrane is $\bar{p}_n\equiv(p_1+p_2)/2=p_2+\Delta p/2$. In practice
$\Delta p\ll (\bar{p}_n-p_t)< \bar{p}_n.$
We  therefore simply replace $\bar{p}_n(s)$ with 
$p_n(s)\approx p_1(s) \approx p_2(s)$ and recast 
(\ref{deltap}) as
\begin{equation}
\label{pnpt}
\gamma\int_{-\infty}^{\infty}{\left[p_n-p_t\right]dy}=
\sigma \gamma + \epsilon \gamma^3  - \epsilon \ddot{\gamma}.
\end{equation}

In the lattice-Boltzmann method~\cite{roth1,doole}, 
the momentum
flux tensor ${\bf \Pi}$ is expressed in terms of the 
discretized 
velocity
distribution functions $n_i({\bf x},t)$
\begin{equation}
\label{momflux}
{\bf \Pi}({\bf x},t)=\sum_i{n_i({\bf x},t){\bf c}_i{\bf c}_i}\,,
\end{equation}
where 
$n_i({\bf x},t)$ represents the positive real-valued 
occupation number at a
given site ${\bf x}$ and time $t$ with velocity 
${\bf c}_i$. Similar expressions hold for the fluid density 
and velocity
\begin{eqnarray}
\rho&=&\sum_i{n_i({\bf x},t)}\\
\rho{\bf u}&=&\sum_i{n_i({\bf x},t){\bf c}_i}\quad .
\end{eqnarray}
In the present work, 
interfaces separate the two components,
called ``red'' and ``blue'', of a 
binary fluid.
The sum of the red occupation numbers
$r_i$ and the blue ones $b_i$ at 
each lattice site is preserved:
\begin{equation}
\label{eq:colors}
n_i({\bf x},t)=r_i({\bf x},t)+b_i({\bf x},t)\,.
\end{equation}
Starting from this distinction in terms of
color attributes, it is possible to build
a numerical method and study
miscible and immiscible fluids, and, after 
suitable generalizations, the physics
of even more complex fluids can be discussed
\cite{roth1,bogho}.

\par The different components of ${\bf \Pi}$ are related to the 
macroscopic pressure tensor. Let us assume, for example,
that the interface is oriented parallel to the
$x$-axis in a neighborhood of a given point $(x_p,0)$. Then
$\Pi_{yy}$ and $\Pi_{xx}$ replace $p_n$ and $p_t$ respectively
in (\ref{pnpt}), resulting in 
\begin{equation}
\label{pipi}
\gamma\!\sum_{k=-\infty}^{\infty}{\sum_i{
n_i(x_p,k,t)\left(c_{iy}^2-c_{ix}^2\right)}}=
\sigma \gamma + \epsilon \gamma^3  - \epsilon \ddot{\gamma}\,,
\end{equation}
where $k$ is the discrete coordinate running along 
the $y$-axis.
\par The microscopic dynamics governing the time evolution
of the distribution functions is described by the
Boltzmann equation. Lattice-Boltzmann models
rely on a discretized form of it,
namely~\cite{roth1,doole,frisch}
\begin{equation}
\label{frischi}
n_i({\bf x}+{\bf c}_i,t+1)=n_i({\bf x},t)+
    \Delta_i[n({\bf x},t)],
\end{equation}
where the last term is the collision operator,
which accounts for the change in the occupation
numbers due to collisions at the lattice sites.
For practical reasons, $\Delta_i[n({\bf x},t)]$
is usually replaced by its linear expansion
around the equilibrium populations 
$n_i^{\mbox{\scriptsize eq}}({\bf x},t)$
\cite{higue}
\begin{equation}
\label{linearize}
\Delta_i[n({\bf x},t)] = 
  \sum_j{L_{ij}\left[n_j({\bf x},t)-
    n_j^{\mbox{\scriptsize eq}}({\bf x},t)\right]} ,
\end{equation}
where $L_{ij}$ is a matrix of costant coefficients.
A further simplification consists in substituting
$L_{ij}$ with a diagonal operator 
$\lambda_B \delta_{ij}$, where $\lambda_{B}$ 
is the relevant eigenvalue of the linearized 
Boltzmann operator, so that (\ref{frischi})
simplifies in the single relaxation time 
lattice-Boltzmann model~\cite{qian,chenchen}
\begin{equation}
\label{boltz}
n_i({\bf x}+{\bf c}_i,t+1)=(1+\lambda_{B})n_i({\bf x},t)-
\lambda_{B} n_i^{\mbox{\scriptsize eq}}({\bf x},t)\,.
\end{equation}
This expression conserves mass and momentum.
Also, the populations $n_i$, in the absence of external 
forcing, converge to the equilibrium occupation
numbers $n_i^{\mbox{\scriptsize eq}}$.

\par Thermal fluctuations are introduced in the model by 
adding a stochastic term $\Delta_i^{\prime}({\bf x},t)$
to the right-hand side of (\ref{boltz}) such that 
\cite{ladd1}
\begin{equation}
\label{stochast}
\Delta_i^{\prime}({\bf x},t) \propto
  \sum_{\alpha,\beta}{\sigma_{\alpha \beta}^{\prime}
      ({\bf x},t)\,\left(c_{i\alpha}c_{i\beta}-
        {c^2}\delta_{\alpha\beta}/D\right)}\,.
\end{equation}
Here $D$ is the dimensionality of the lattice and
the random fluctuations $\sigma_{\alpha \beta}^{\prime}$
are uncorrelated in space and time~\cite{land2}, and 
sampled from a Gaussian distribution such that
$$
  \left\langle\sigma_{\alpha \beta}^{\prime}({\bf x},t)
   \sigma_{\eta \zeta}^{\prime}({\bf x}^{\prime},t^{\prime})
    \right\rangle = 
$$
\begin{equation}
\label{gauscorfun}
 A\,\delta_{{\bf x}{\bf x}^{\prime}}\delta_{tt^{\prime}}
  \left(\delta_{\alpha\eta}\delta_{\beta\zeta}+
        \delta_{\alpha\zeta}\delta_{\beta\eta}-
         \frac{2}{3}\delta_{\alpha\beta}\delta_{\eta\zeta}\right)
\end{equation}
where the variance $A$ is related to the effective temperature $T$
of the fluid,
\begin{equation}
\label{temperature}
 A = 2\rho\nu k_BT\lambda_B^2\,,
\end{equation}
by means of the
fluctuation-dissipation theorem~\cite{ladd1}. 

In order to reproduce the desired surface tension
and bending stiffness of the interface, so that
the left-hand side of (\ref{pipi}) does not
vanish in the proximity of the boundary, 
one may suitably 
perturb the single relaxation time 
lattice-Boltzmann model by adding a term 
$\Delta_i^{\prime\prime}$ to the right-hand side.
Our perturbation of (\ref{boltz}) reads
\begin{equation}
\label{pert}
\Delta_i^{\prime\prime} \equiv
\left(S+E\gamma^2-E{\ddot{\gamma}}/{\gamma}\right)
\left|{\bf f}\right|
\sum_{\alpha \beta}{\left(c_{i\alpha}c_{i\beta}-
{c^2}\delta_{\alpha\beta}/D\right)
\frac{\mbox{f}_{\alpha}\mbox{f}_{\beta}}{\mbox{f}^2}}
\end{equation}
where $S$ and $E$ are two adjustable parameters
corresponding to the physical surface tension $\sigma$
and bending rigidity $\epsilon$, and 
${\bf f}({\bf x},t)$ is the local color gradient~\cite{roth1}
as defined in Appendix~\ref{appendixb}.
In order to measure the geometric properties of the
membrane---the curvature $\gamma$ and its 
derivatives---appearing in equation (\ref{pert}),
we adopt an explicit procedure of localization
of the different connected components of the interface.
The method is described in some detail in 
Appendix~\ref{appendixb}. There we show how, 
for each component, we map the boundary piecewise 
to polynomials, from which we extract the curvature 
and its derivatives.
This process does not result in a significant
time consumption for the whole simulation, as it
is performed only once at the beginning of each 
time step and it is limited to the lattice 
sites which constitute the interface.

\par Expression (\ref{pert}) preserves the total mass 
and momentum at a given lattice site, as one can
verify by recalling the identities
\begin{eqnarray*}
\sum_i{c_{i\alpha}} &=& 0\\
\sum_i{c_{i\alpha}c_{i\beta}} &=& 
{b_m c^2}\delta_{\alpha\beta}/D\\
\sum_i{c_{i\alpha}c_{i\beta}c_{i\gamma}} &=& 0,
\end{eqnarray*}
which hold true for tensors with hypercubic 
symmetry.
By generalizing the argument set forth in
Section 10.2 of~\cite{roth1}, 
it can be shown
that the inclusion of 
(\ref{pert}) into the right-hand side of (\ref{boltz}) 
generates a surface 
tension $\sigma$ and a bending rigidity $\epsilon$
of the interface, which are related to $S$  and $E$ 
respectively through the linear relation 
\begin{equation}
\label{prop}
\frac{\sigma}{S}=\frac{\epsilon}{E}=
-\frac{192 \rho}{\lambda_{B}}
\end{equation}
valid for the face-centered  
hypercubic lattice (FCHC), which we shall employ
throughout this work.
The above result follows from 
replacing equations (10.14) and (10.3) 
in~\cite{roth1} with our
equations (\ref{pipi}) and the result of the
perturbation of (\ref{boltz}) by (\ref{pert}) 
and carrying out an analysis similar to that 
one in~\cite{roth1}.

\section{Testing the model}\label{test}

The surface tension parameter $S$ is set to 
zero so that purely elastic effects can be
studied. In section \ref{bubb}, the results
for a spherical bubble  surrounded by
fluids on both sides are presented. 
In section \ref{bend}, the bending wave
dispersion relation is discussed by studying
the damped oscillations of
sinusoidal interfaces.

\subsection{Spherical bubbles}
\label{bubb}

\begin{figure}[htb!]
  \begin{center}
    \epsfig{file=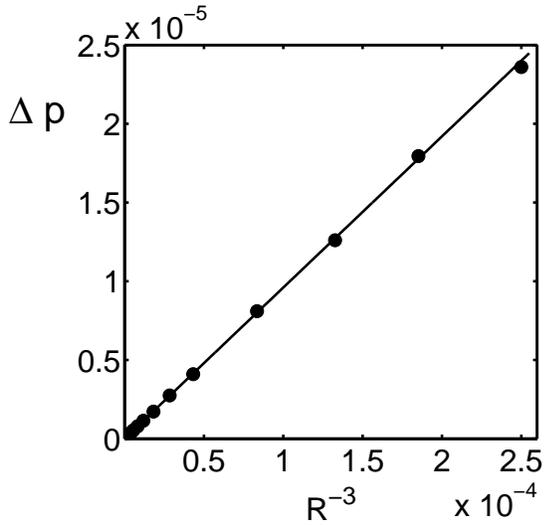,width=.4\textwidth}
    \caption{Verification of the
             $\Delta p=\epsilon/R^3$ law
             for a fluid bubble immersed in an 
             immiscible sea. The solid line is the 
             theoretical prediction for the parameters
             specified in the text. 
             The pressure gap $\Delta p$ is determined from the equation
             of state $p(\rho) = {3\rho}/10$, where 
             $\rho$ is the fluid density.
             The radius $R$ of the bubble is expressed in 
             lattice units.}
    \label{elastbubbl}
  \end{center}
\end{figure}

The lattice-Boltzmann simulation is initialized as
a spherical bubble of a fluid, here called 
``red'' for practical purposes, immersed in a bath
of ``blue'' fluid with linear dimension $L$ much 
larger than the radius $R$ of the sphere.
The pressure gap $\Delta p$ between the red inner part
of the bubble and the outside blue sea is predicted
by replacing (\ref{prop}) in (\ref{deltap}), resulting in
\begin{equation}
\label{lapl}
\Delta p = 
-\frac{192 \rho E}{\lambda_{B}} \gamma^3 =
-\frac{192 \rho E}{\lambda_{B} R^3}
\end{equation}
as $\sigma$ is set to vanish in our experiments,
$\gamma=1/R$ and $\ddot{\gamma} = 0$ for
a sphere. The pressure is computed by 
measuring the fluid density according to
the equation of state (in the absence of 
a net momentum)~\cite{roth1}
\begin{equation}
\label{pres}
p(\rho) = \frac{b_m \rho}{2 b},
\end{equation}
where $b_m$ is the number of velocity 
vectors (24 for FCHC), while $b\equiv 
b_m+b_r$ includes the number $b_r$ 
of rest particles (in this case $b_r = 16$).
Simulations were performed for different
bubble sizes, ranging from $R=8$ to $R=64$,
with average density of $0.5$ particles
per lattice site, $\lambda_{B} = -1$ and $E=10^{-3}$.
The measured values of $\Delta p$ are plotted  versus $R^{-3}$
in Figure \ref{elastbubbl}.
The agreement with the predicted relation
$\Delta p = 0.096 R^{-3}$,
drawn as a solid line, is very good.
For bubble radii smaller than eight lattice 
units, discretization effects cause the measured
curvature to be off more than $15\%$, 
and ultimately wrong when the radius 
of curvature is of the order of the interface thickness,
that is about four lattice units.

\subsection{Bending waves}\label{bend}

\begin{figure}[htb!]
  \begin{center}
    \epsfig{file=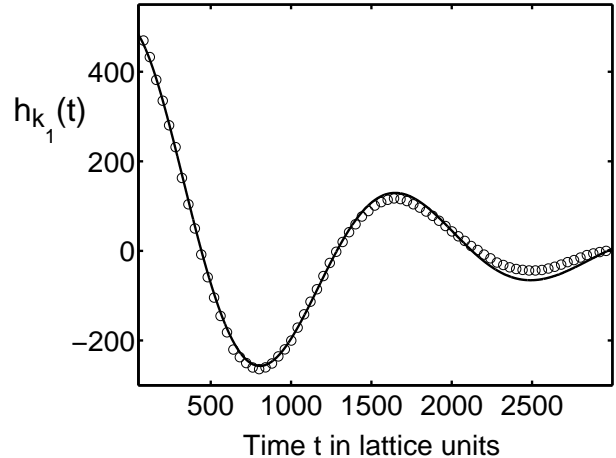,width=.45\textwidth}
    \caption{We report in figure the results of the 
             lattice{-}Boltzmann simulation 
             of a bending wave with wavelength $\lambda = 100$
             lattice units (circles). 
             The theoretical predictions from the normal-mode 
             analysis, equations (\ref{mode}) and (\ref{disp}),
             are graphed as a solid line. $h_{k_1}(t)$ is the 
             Fourier transform in lattice units of the interface
             profile, corresponding to the wavenumber 
             $k_1 = 2\pi/\lambda$.}
  \label{eldampes}
  \end{center}
\end{figure}

While bubble pressure gaps verify the equilibrium
properties of the membrane, the simulation of
bending waves provide an effective tool for
testing the dynamics of the fluid-interface
coupling.
In testing the bending wave dispersion relation, we
initialize the system as a square region filled with
two immiscible fluids with the same densities and
viscosities. Such fluids are separated by a
sinusoidal interface, whose wavelength $\lambda$ 
is equal to the linear dimension of the box.
We impose periodicity along the
horizontal axis, while free-slip boundary conditions are 
prescribed along the vertical axis.
The damped oscillations of bending waves for
membranes immersed in viscous fluids have been analytically
studied in the literature~\cite{mors}. For the initial conditions 
we impose one expects the time-dependence of the first
normal mode  to be described by
\begin{equation}
\label{mode}
h_{k_1}(t)=h_{k_1}^0\cos[\Re(\omega)t]\, e^{-\Im(\omega)t},
\end{equation}
where $k_1=2\pi/L$, with $L$ the size of the box,
and $h_{k_1}^0$
is the initial amplitude of the sinusoidal wave.
The complex angular frequency $\omega$ is related 
to the wave number $k\equiv2\pi/\lambda$ through the dispersion 
relation~\cite{broc}
\begin{equation}
\label{disp}
\begin{array}{cl}
 {\displaystyle \omega^2=\frac{ \epsilon k^5}{2\rho}
     \left(1-{k}/{q}\right)}\\
   q=\sqrt{k^2-i{\omega}/{\nu}}
\end{array}
\end{equation}
where $\nu=\mu/\rho$ is the kynematic shear viscosity.
The numerical simulation of a bending wave for a system
defined by 
$E=0.25, L=100, \rho=0.5, \lambda_{B} = -1.0, h^0_{k_1}/L=0.05$
is given as an example in Figure \ref{eldampes}.
By means of (\ref{prop}), (\ref{disp}) and the relation 
between the viscosity $\nu$ and the Boltzmann eigenvalue 
$\lambda_{B}$~\cite{roth1},
\begin{equation}
\label{visc}
\nu = -\frac{1}{3}\left(\frac{1}{\lambda_{B}}
+\frac{1}{2}\right).
\end{equation}
Therefore one may fully predict 
the behavior of the normal mode (\ref{mode}).
$h_{k_1}(t)$ was computed as the Fourier transform of the
experimental interface profile $h(x,t)$, recorded at
each time step. 
Figure \ref{eldampes} shows the 
evolution of $h_{k_1}(t)$ according
to (\ref{mode}) (solid line) and our numerical results
(circles).
By fitting the first cycle of the time evolution of 
$h_{k_1}(t)$ to a curve of the form (\ref{mode})
($h_{k_1}^0$ is initially prescribed) we collected
the numerical data about the complex angular
frequency $\omega$, whose real and imaginary parts
correspond to the oscillation frequency and
damping rate, respectively, of the bending wave.
We repeated the simulation of Figure \ref{eldampes}
for different wavelengths,
collecting the data about the damping rates $\Im(\omega)$
and the oscillation frequencies $\Re(\omega)$.
\begin{figure}[htb!]
  \begin{center}
    \epsfig{file=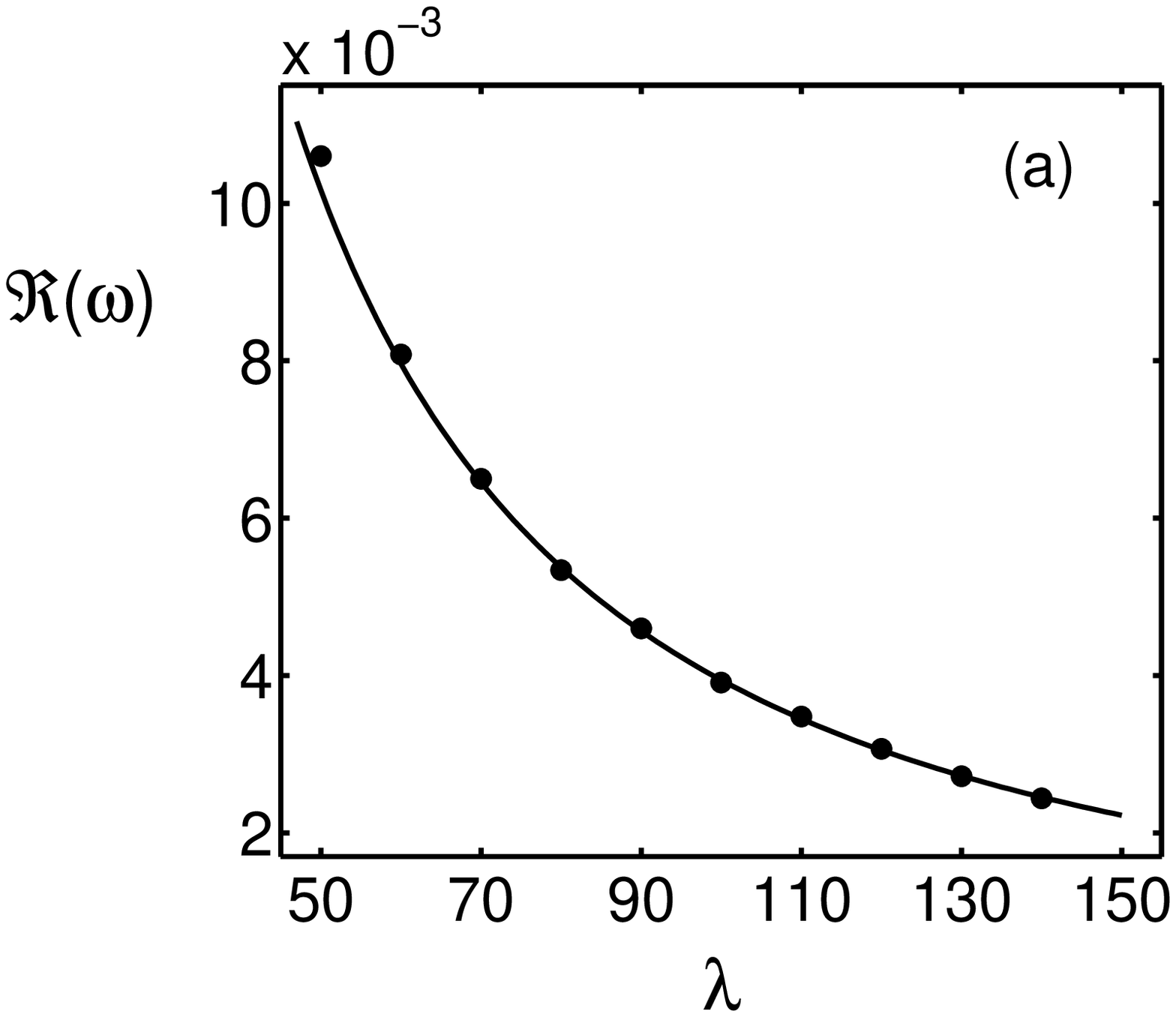,width=.4\textwidth}
    \epsfig{file=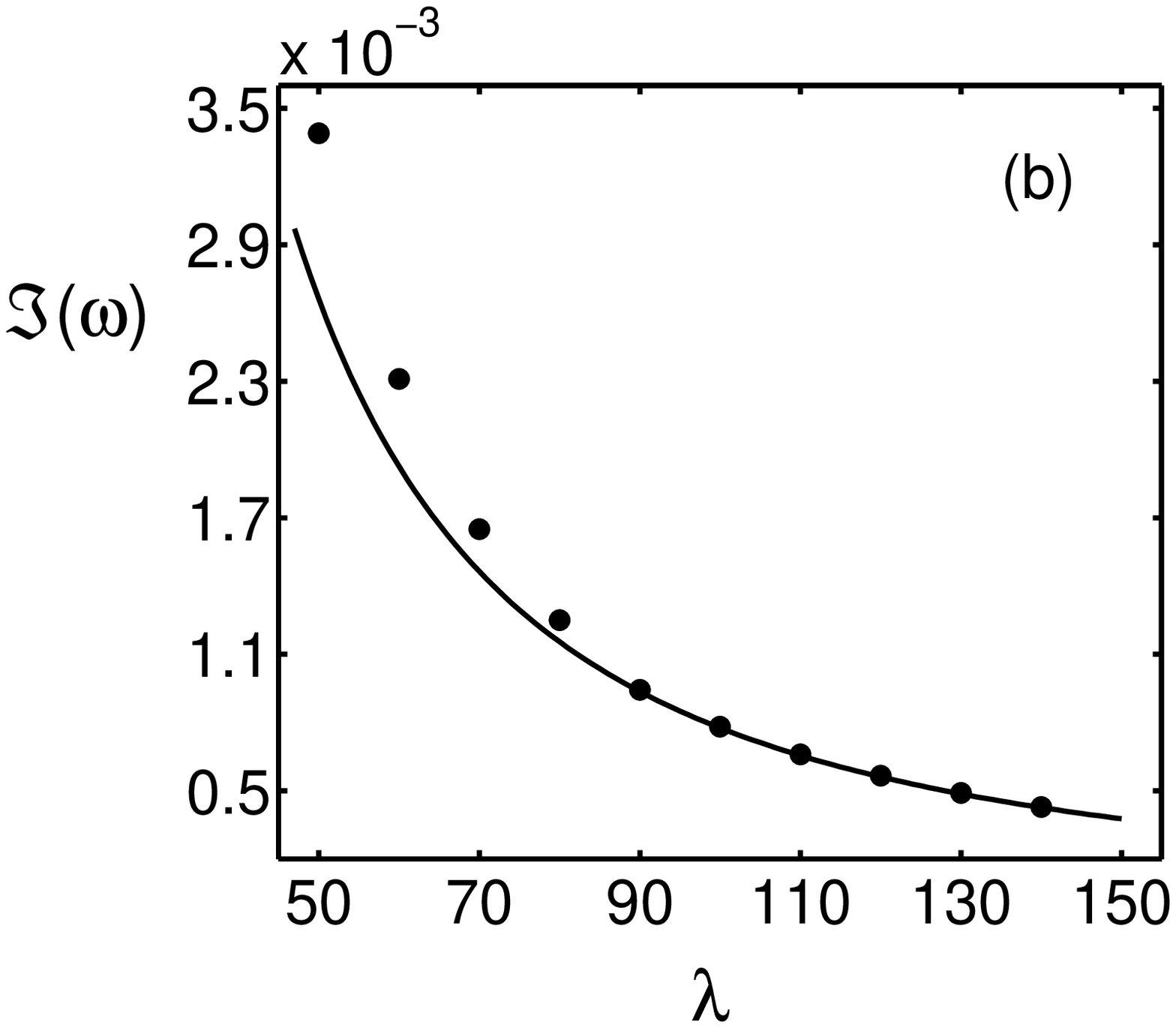,width=.4\textwidth}
    \caption{(a): Oscillation frequencies collected 
             from the simulation
             of bending waves with wavelengths ranging
             from $50$ to $140$ lattice units are 
             represented with filled circles. The
             numerical solution of (\ref{disp})
             is drawn as a solid line. 
             (b): Damping rates for the 
             same experiment correspond to the 
             imaginary part of the complex 
             frequency $\Im(\omega)$. The discrepancy
             at shorter wavelenghts is due to the damping
             action
             of the effective surface tension discussed 
             in Section \ref{numeraccu}.}
    \label{elastdamp}
  \end{center}
\end{figure}
In Figure \ref{elastdamp} we produce the experimental results
together with the numerical solution of (\ref{disp}).
We notice that the agreement between experimental data
and theoretical predictions improves for larger systems 
and the relative errors reduce to few percent for 
wavelengths of ${\cal O}(10^2)$ lattice units.
The relative error is here defined in terms of the
theoretical and experimental time evolution of the 
damped wave as
\begin{equation}
\varepsilon(\lambda)=
\sqrt{{1 \over \tau}\sum_{t=0}^{\tau}{
     \left[1-{h_{k_1,ex}(t)/
     h_{k_1,th}(t)}\right]^2}}\;,
\end{equation}
where $\tau$ is twice the oscillation period of the bending
wave.
Table \ref{relerr} 
\begin{table}[htb!]
 \protect
  \caption{Relative errors $\varepsilon(\lambda)$
           for the sinusoidal
           bending waves discussed in Section \ref{bend} and 
           in Figure \ref{elastdamp}.}
    \begin{tabular}{cc} 
      $\lambda$&$\varepsilon(\lambda)$	
                     \\ \hline
      50 	&0.2741\\ 
      60 	&0.1915\\ 
      70	&0.1262\\ 
      80	&0.0833\\ 
      90	&0.0640\\ 
      100	&0.0519\\ 
      110	&0.0428\\ 
      120	&0.0371\\ 
      130	&0.0323\\ 
      140	&0.0289\\ 
    \end{tabular}
  \label{relerr}
\end{table}
reports the relative errors for different 
wavelengths, confirming their fast convergence 
towards ${\cal O}(10^{-2})$ when $\lambda>70$.
The error at small wavelengths is due to a numerical 
artifact that manifests itself as an effective surface
tension. We postpone further discussion of this 
subject to Section \ref{numeraccu}.

\section{Thermal fluctuations of elastic interfaces: Theory}

We now turn to a study of fluctuating membranes.
This section introduces the theoretical results relevant
to our discussion. A comparison with numerical experiments
will then be presented in succeeding sections.
The fluctuation-dissipation theorem provides us with a 
unified description of the steady state and
the non-equilibrium growth (roughening) 
of fluctuating interfaces.
Emphasis is given to the frequency power-spectrum, as
it conveys all the relevant information about the decay 
of the correlation function of the fluctuations. 

\subsection{Correlation functions from the 
            fluctuation-dissipation theorem}

A detailed discussion of the fluctuation-dissipation theorem and
fluctuating hydrodynamic interfaces driven by surface tension
has been presented in~\cite{roth3,roth4}.
Here we summarize the theoretical results 
for 1D interfaces with 
bending stiffness.

The interface height will be denoted by $h(x,t)$ and its 
Fourier transform in wave-number space by $h_k(t)$.
We also introduce the Fourier transform in time as
\begin{equation}
h_k(\omega)= \frac{1}{\Theta}\int_0^
        {\Theta}{dt \,h_k(t)\, e^{-i\omega t}}\,,
\label{hkomega}
\end{equation}
where $\Theta$ is the size of the time integration domain.
When $\Theta \rightarrow \infty$ the 
integral in (\ref{hkomega}) will be replaced 
by $\frac{1}{\Theta}\int_0^{\Theta}\rightarrow 
\frac{1}{2\pi}\int$.

It can be shown~\cite{roth4,herp} that the frequency 
power-spectrum can be written in terms of the 
response function $\Gamma_k(\omega)$ as
\begin{equation}
\left|h_k(\omega)\right|^2= \frac{\Theta}{2\pi^2}
      \frac{k_B T}{\omega^2 L}
         \,\Re\,[\Gamma_k(\omega)]\,,
\label{fps}
\end{equation}
where
\begin{equation}
\Gamma_k^{-1}(\omega)=\frac{2\omega\rho}
    {ik\left(1-{k}/{q}\right)} -
    \frac{\epsilon k^4}{i\omega}
\end{equation}
and $q$ is given in (\ref{disp}).
The frequency power-spectrum carries information about
the fluctuation correlation functions. Indeed, by
means of the Wiener-Khintchine relation, (\ref{fps}) can
be used to show that
\begin{equation}
\left\langle h_k(t)h_k^*(0)\right\rangle =
  \frac{k_BT}{2\pi L}\int_{-\infty}^{\infty}
    {\frac{d\omega}{\omega^2}\, e^{-i\omega|t|}\,
      \Gamma_k(\omega)}\,.
\label{flucorfu}
\end{equation}
\begin{figure}[htb!]
  \begin{center}
    \epsfig{file=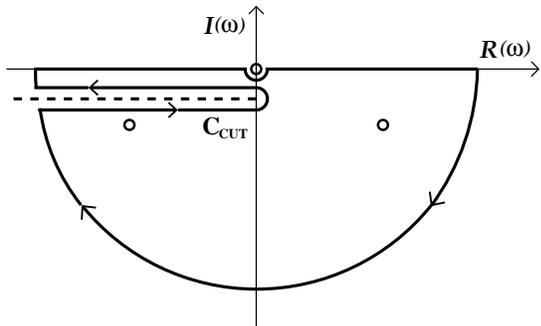,width=.4\textwidth}
    \caption{Integration contour for the fluctuation 
             correlation function (\ref{flucorfu}). 
             The poles in the lower half-plane 
             corresponds to the bending wave contribution,
             while the pole at the origin represents the 
             steady state power-spectrum.}
    \label{contour}
  \end{center}
\end{figure}
The integration in (\ref{flucorfu}) can be performed 
in the complex $\omega$-plane by means of the contour drawn 
in Figure \ref{contour}, so that 
\begin{equation}
\left\langle h_k(t)h_k^*(0)\right\rangle=
  \left\langle h_k(t)h_k^*(0)\right\rangle_{\rm bw}+
  \left\langle h_k(t)h_k^*(0)\right\rangle_{\rm cut}
\end{equation}
is the sum of the contributions from the two poles
(bending wave) and from the branch cut.
In the case of oscillating bending waves the 
contribution
from the two poles results in
\begin{eqnarray}
&\left\langle h_k(t)h_k^*(0)
\right\rangle_{\rm bw}=\nonumber\\
&{\displaystyle \frac{k_BT}{\epsilon k^4 L}}
\left[\cos|\Re(\omega)t| - \left|
  \frac{\Im(\omega)}{\Re(\omega)}\right|
    \sin|\Re(\omega) t|\right]\,
       e^{-|\Im(\omega) t|}\:,
\end{eqnarray}
where $\omega$ as a function of $k$ 
is given by the dispersion relation (\ref{disp}).
Also, in the long time limit, the contribution
from the branch cut reads
\begin{equation}
\left\langle h_k(t)h_k^*(0)\right\rangle_{\rm cut}=
\frac{2k_BT\rho\nu^{1/2}}{\epsilon^2k^8L}\:
     \frac{e^{-k^2\nu|t|}}{|t|^{3/2}}\,.
\label{branchcut}
\end{equation}
One can show by means of the fluctuation-dissipation   
theorem~\cite{roth4} that the correlations among 
the fluctuating
forces acting on the interface, and due to the 
uncorrelated thermal excitations in the bulk, have 
indeed
the same temporal decay as (\ref{branchcut}).
A special case is the one of $t=0$ in (\ref{flucorfu}).
The integration contour can then be closed in the 
upper 
half-plane, including only the pole at $\omega=0$. As a
result
\begin{equation}
\left\langle\left|h_k(0)\right|^2\right\rangle=
\frac{k_BT}{\epsilon k^4L}
\label{steadystate}
\end{equation}
represents the mean square amplitude of the 
$k$-th mode~\cite{broc}, which can also be understood
by applying the energy equipartition theorem to the 
free energy (\ref{freen}).

\subsection{Interface roughening}
\label{introu}

The mean-square width $W^2(L,t)$ of an interface with
vanishing mean height, defined as
\begin{equation}
 W^2(L,t)\equiv\frac{1}{L}\int_0^L{dx\, h^2(x,t)}
\label{width}
\end{equation}
will here be used to describe the interface roughening.
By using Parseval's relation, (\ref{width}) can
be recast as
$W^2(L,t)=\sum_k{\left|h_k(t)\right|^2}$.
At the steady state, the average width can be evaluated
by means of (\ref{steadystate}),
\begin{equation}
 W^2(L) = \frac{2k_BT}{\epsilon L}\,
   \sum_{n=1}^{\infty}\left(\frac{L}{2\pi n}\right)^{\!4}=
     \frac{k_BT}{720\,\epsilon}\,L^3\,,
\label{steadywidth}
\end{equation}
where the factor 2 accounts for the two possible
orientations (for a 1-dimensional interface) of the wave vectors.
This result shows that at equilibrium $W/L\propto L^{1/2}$,
that is the relative width apparently increases 
indefinitely with the square root
of the linear size of the computational box. 

\par In reality, the argument given above and culminating
with equations (\ref{steadystate}) and (\ref{steadywidth})
implicitly approximate the free energy 
${\cal F}=\frac{1}{2}\,\epsilon\int{ds\, \gamma^2}$
with the more convenient expression
${\cal F}_{\rm ap}=\frac{1}{2}\,\epsilon\int{dx\, (d^2h/dx^2)^2}$.
Such approximation is certainly plausible when 
$\gamma \approx d^2h/dx^2$, that is when the interface-width/length
ratio is small. In practice, however, 
when $W/L > 1/2$ the local
curvature is significantly different from its linearization
$d^2h/dx^2$, and the original expression (\ref{freen})
for the free energy should be considered.
Therefore (\ref{steadystate}) and (\ref{steadywidth})
hold true only for small values of $W/L$, while
when $W/L > 1/2$ nonlinear terms in the curvature and
consequently in the dynamics prevent the interface-width/length 
ratio from growing indefinitely with the size of the system.
On the other hand, the lattice-Boltzmann method we are 
currently describing  
does not suffer from such a limitation. Indeed, the interface 
dynamics discussed in Section \ref{interdyn} originates 
in our model from
the perturbation (\ref{pert}), which corresponds to the exact 
analytical form of the elastic force 
$F_{\rm el}=-\epsilon \,d^2\gamma/ds^2$ and not to its linearized 
approximation $F_{\rm el}\approx-\epsilon \,d^4h/dx^4$,
deducible from  ${\cal F}_{\rm ap}$.
This allows 
us to study and simulate the dynamics of membranes
in the nonlinear regime.

In order to analyze the non-equilibrium roughening of 
the interface, two immiscible fluids at the same
temperature $T$ are brought in contact at time $t=0$.
The interface between them is initially 
flat. Due to the thermal fluctuations in the 
bulk, standing bending waves will be excited on
the interface with frequencies given by the dispersion 
relation
\begin{equation}
 \omega_o(k)=\sqrt{\frac{\epsilon}{2\rho}}\,k^{5/2}\,,
\end{equation}
where for simplicity we consider the inviscid limit
($\nu=0$) of (\ref{disp}).
In analogy with the argument of~\cite{roth3,roth4}
and by means of the fluctuation-dissipation theorem,
one anticipates that the power spectrum is given by
\begin{equation}
\left\langle\left|h_k(t)\right|^2\right\rangle=
  \frac{2k_BT}{\epsilon L k^4}\, \sin^2 [\omega_o(k)t]\,,
\label{growingphases}
\end{equation}
so that the mean-square width of the interface is expected to 
grow as
\begin{eqnarray}
&W^2(L,t)=\nonumber\\
&{\displaystyle \frac{4k_BT}{\epsilon L}\,
   \sum_n{\left(\frac{L}{2\pi n}\right)^{\!4}\,
     \sin^2\left[\sqrt{\frac{\epsilon}{2\rho}}
        \left(\frac{2\pi n}{L}\right)^{\!5/2}\,t\right]}}\:,
\label{intergro}
\end{eqnarray}
where the thermal average of $W^2$ is understood.
The inviscid form of $W^2(L,t)$ as given by 
(\ref{intergro}) does not hold for long times, since it does
not relax to the equilibrium (\ref{steadywidth}) as it 
should if the effects of a finite viscosity were taken into account.
Nonetheless, the initial excitations are all in phase, as 
the interface starts with a flat profile, so (\ref{intergro})
is a correct description of the short-time non-equilibrium growth 
of the interface width.

\par From (\ref{steadywidth}) and (\ref{intergro}) one finds that
$W^2$ scales according to
\begin{equation}
W^2(L,t)=L^3 \,g\!\left(t/L^{5/2}\right)\,,
\end{equation}
where for large times
\begin{equation}
\lim_{t \rightarrow \infty}{
      g\!\left(t/L^{5/2}\right)}=
         \frac{k_BT}{720\,\epsilon}\:,
\end{equation}
while the short-time limit is given by
\begin{equation}
g\!\left(t/L^{5/2}\right)= 
  \frac{k_BT}{4\,\epsilon \pi^4}\,
    \sum_n{\frac{1}{n^4}\,\sin^2\!
     \left[\sqrt{\frac{\epsilon}{2\rho}}
       \left(\frac{2\pi n}{L}\right)^{\!5/2} t\right]}\,.
\end{equation}
One can obtain an analytical form for the above expression 
by approximating the summation with an integration, so that
(\ref{intergro}) becomes
\begin{equation}
W^2(t)=\frac{4k_BT}{5\pi(2\rho)^{3/5}\,\epsilon^{2/5}}\;
    t^{6/5}\!\int_0^{\infty}{dx\,\frac{\sin^2(x)}{x^{11/5}}}\;,
\end{equation}
where we set $x=\omega_o(k)t$. The integral evaluates 
to~\cite{grads}
\begin{equation}
\int_0^{\infty}{dx\,\frac{\sin^2(x)}{x^{11/5}}}=
  \frac{5^2\,2^{1/5}}{6}\cos(2\pi/5)\,
    \Gamma(4/5)=1.7219\dots
\end{equation}
and the roughening of the interface thus scales according to
\begin{equation}
W^2(t)=\frac{2^{3/5}\,5\cos(2\pi/5)\, \Gamma(4/5)}
    {3\pi}\,\frac{k_BT}{\rho^{3/5}\,\epsilon^{2/5}}\:t^{6/5}\:.
\label{earlyscaling}
\end{equation}
We expect (\ref{earlyscaling}) to describe the growth of the 
interface until the crossover time in which equilibrium
is attained.
The crossover time $t_c$ may be approximated by the time necessary 
for the longest wavelength excitation to reach its
maximum amplitude, that is $t_c=T/4$. If the dispersion relation
(\ref{disp}) is used to estimate the period $T$, we conclude 
that 
\begin{equation}
t_c=\frac{\pi}{2\,\omega_o}=\frac{1}{8}
       \sqrt{\frac{\rho L^5}{\pi^3\epsilon}}\:.
\label{crossovertime}
\end{equation}
We shall compare the lattice-Boltzmann numerical simulations
with the theoretical results of this section
in the following.

\section{Thermal Fluctuations of elastic interfaces: Simulations}
\label{thermy}

In this section we present the results of the simulations 
performed using the lattice-Boltzmann method with a
9-velocity square lattice (that is, an FCHC lattice projected
in 2-dimensions)~\cite{roth1,doole}. The 
computational box of size $L$x$L$ is filled with two immiscible
fluids pre-thermalized to a common temperature and separated 
by a thin interface. The two fluids have the 
same densities and viscosities. We study the dynamical effects
of thermal fluctuations at equilibrium and out of equilibrium,
when the interface grows up to the steady
state given by (\ref{steadywidth}). The interfaces we consider 
are driven by the coupling with the surrounding fluids and 
by purely elastic forces, tuned by the bending modulus
$\epsilon$, as defined in Section \ref{interdyn}.

\subsection{Steady state}\label{section:steady}

\begin{figure}[htb!]
  \begin{center}
    \epsfig{file=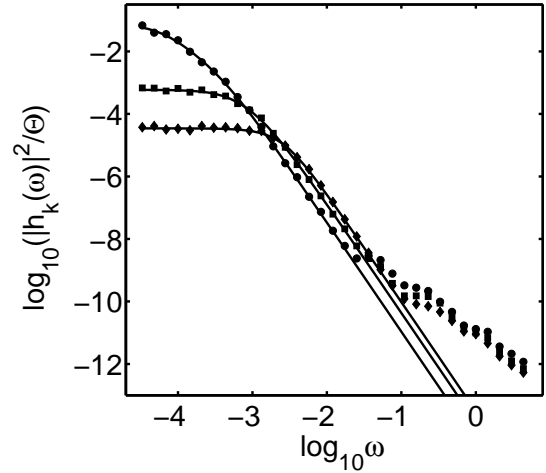,width=.4\textwidth}
    \caption{Frequency power-spectrum for a fluctuating 
             interface of size $L=64$, whose time evolution
             was monitored for
             $2^{20}$ time steps. Experimental data for
             wavenumbers $k=2\pi/L, 4\pi/L$ and $6\pi/L$
             are reported as filled circles, squares and
             diamonds respectively. The solid lines represent 
             the theoretical predictions from (\ref{fps}), 
             showing that the agreement spans about three
             orders of magnitude.}
    \label{fig:freqpowspec}
  \end{center}
\end{figure}

The evolution of an interface of size $L=64$ was monitored
for $2^{20}$ time steps. The resulting log-log plot of the 
frequency power-spectrum is given in Figure \ref{fig:freqpowspec}.
We distinguish three regions:

$\bullet$ A plateau corresponding to the low-frequency limit 
of (\ref{fps}),
\begin{equation}
\lim_{\omega \rightarrow 0}{\left|h_k(\omega)\right|^2/\Theta}=
\frac{2\rho k \nu}{\pi^2}\,\frac{k_BT}
   {\left(\epsilon k^4\right)^2L}\,,
\label{lowlimit}
\end{equation}
that shows explicitly a dependence on the bending stiffness
$\epsilon$~\cite{foot1}.
Low-frequency excitations are thus controlled by elastic forces 
and viscous effects. In figure this region spans a frequency interval
of about two orders of magnitude, up to where
$\omega$ approaches the pole of the response function $\Gamma$.

$\bullet$ A transition region in proximity of 
the bending wave frequency 
given by the dispersion 
relation (\ref{disp}), a pole for (\ref{fps}). Notice that 
a peak at such frequency manifests itself at higher wavenumbers. 
This behavior is
opposite to the one observed for interfaces driven by 
surface tension~\cite{roth3,roth4}.

$\bullet$ The high-frequency region, expected from (\ref{fps}) to 
represent a $\omega^{-7/2}$ decay, independent of either
the bending rigidity or the surface tension of the interface,
according to
\begin{equation}
\lim_{\omega \rightarrow \infty}
     {\left|h_k(\omega)\right|^2/\Theta}=
\frac{\sqrt{2\nu}\,k^2\,k_BT}{8\pi^2 L\,\rho}\,\omega^{-7/2}\,.
\label{highlimit}
\end{equation}
The relative amplitude of this segment of the 
power spectrum is thus determined by viscosity effects alone.

The experimental results reproduced in Figure \ref{fig:freqpowspec}
agree with the theoretical predictions, here drawn as solid lines, over
a range of about three orders of magnitude, including part 
of the high-frequency region. The discrepancy observed at higher 
frequencies is probably due to standing sound waves that create 
a distortion 
in the interface profile~\cite{roth4}.
Since the theoretical prediction (\ref{fps}) has been confirmed 
by our numerical results, the subsequent conclusions, equations
(\ref{steadystate}) and (\ref{steadywidth}), describe the
average wavenumber power spectrum and 
the functional dependence of the interface
width on the linear dimension  of the box. In particular,
from (\ref{steadywidth}) we estimate that $W/L\approx 1/16$ 
for the case just discussed.

\subsection{Effective surface tension}
\label{numeraccu}

More simulations were performed to verify the agreement with the 
theoretically predictable wavenumber power spectrum. 
Different temperatures, 
lattice sizes and bending rigidities were prescribed to
the system. 
\begin{figure}[htb!]
  \begin{center}
    \epsfig{file=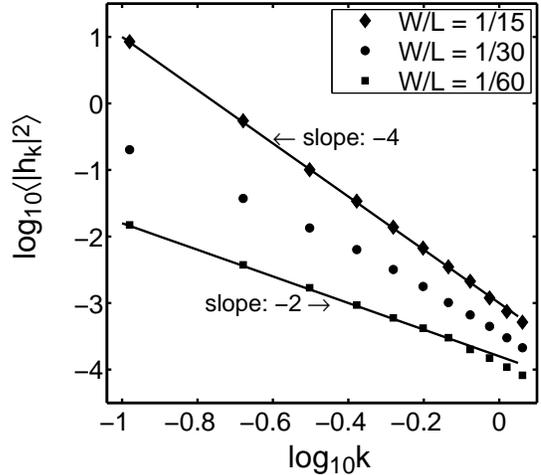,width=.4\textwidth}
    \caption{Power-spectra of fluctuating membranes
             of size $L=60$ lattice units and different bending
             stiffnesses. The interface-width/length 
             ratio, $W/L$, ranges from 1/60 to 1/15.
             When $W/L>1/20$ the purely elastic functional 
             dependence of the power spectrum (with
             slope of $-4$ in a log-log plot) is recovered.}
    \label{fig:steady}
  \end{center}
\end{figure}
In Figure \ref{fig:steady} we report a study of
equilibrium power spectra for a lattice of size $L=60$ and
bending modulus $\epsilon$ ranging from 0.0032 to 0.512. 
With this
set of parameters the ratio $W/L$ spans an interval from 
$1/6$ to $1/60$. We notice a very good agreement
with (\ref{steadystate}) when $W/L>1/20$, but for smaller
ratios the relation describing the steady state power
spectrum is approximated by
\begin{equation}
\left\langle\left|h_k\right|^2\right\rangle=
\frac{k_BT}{\left(\sigma_{\rm eff}k^2 +
     \epsilon k^4\right)L}\,,
\label{steadyeff}
\end{equation}
showing the presence of an effective surface tension 
$\sigma_{\rm eff}$.
By considering the small-$k$ limit of (\ref{steadyeff}) when
$W/L<1/20$, we measured $\sigma_{\rm eff}$ for different points in
the 3-dimensional $\epsilon-L-T$ space. The following empirical scaling
law resulted from our simulations:
\begin{equation}
\sigma_{\rm eff}=
   \frac{1}{L^2}\,{\cal D}\!\left[(W/L)^2\right]=
   \frac{1}{L^2}\,{\cal D}(TL/\epsilon)\,,
\label{effsurtescal}
\end{equation}
where ${\cal D}$ is a scaling function.
\begin{figure}[htb!]
  \begin{center}
    \epsfig{file=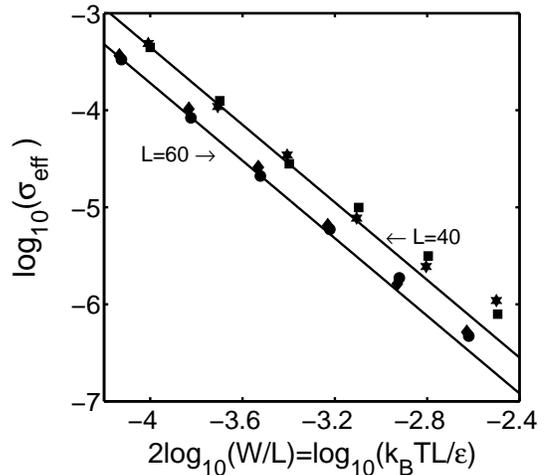,width=.4\textwidth}
    \caption{Effective surface tension as results from the
             power spectra of the simulations of fluctuating
             membranes. Computational boxes with $L=40$
             and 60, different temperatures and bending
             rigidities were considered. We report in
             figure a log-log plot of the data for 
             $\sigma_{\rm eff}$ versus $k_BTL/\epsilon=W^2/L^2$.}
    \label{fig:surfef}
  \end{center}
\end{figure}
Figure \ref{fig:surfef} shows a log-log plot of the
measured values of $\sigma_{\rm eff}$ versus $(W/L)^2$.
$\sigma_{\rm eff}$ decays exponentially for increasing values
of $W/L$; also, when $W/L>1/20$ it practically vanishes and the 
purely elastic dynamics takes place.

In order to discuss the origin of equation 
(\ref{effsurtescal}), we investigate here the 
scaling properties of the ratio 
$\ddot{\gamma}/\gamma$,
the crucial factor in the perturbation (\ref{pert})
of the Boltzmann equation. 
For a 
boundary of the form $h(x)$,
the derivative with respect to
the arc length $s$ can be expressed in terms of 
$x$ as
\begin{equation}
\frac{d}{ds}=\frac{1}{\sqrt{1+{h^{\prime}(x)}^2}}
           \,  \frac{d}{dx}\:,
\end{equation}
where the prime stands for $d/dx$.
From the expression for the curvature,
\begin{equation}
\gamma(x)=\frac{{h^{\prime\prime}(x)}}
        {\left(1+{h^{\prime 2}}\right)^{3/2}}\,,
\label{curvetta}
\end{equation}
one obtains
\begin{equation}
\frac{\ddot{\gamma}}{\gamma}=\frac{h^{(4)}}
      {h^{\prime \prime}{(1+{h^{\prime 2}})}}-
      10\frac{h^{\prime}h^{\prime\prime\prime}}
       {(1+{h^{\prime 2}})^2}-
       3\frac{(1-5{h^{\prime 2}})
       {h^{\prime\prime}}^2}{(1+{h^{\prime 2}})^3}
\label{appb:exaratio}\, .
\end{equation}
In the case of  sinusoidal bending waves
and fluctuating interfaces with
horizontal periodicity as a boundary
condition, the profile $h(x)$ satisfies~\cite{stel}
\begin{equation}
h^{(n)}(x)=
   \frac{W}{L^n}\,{\cal T}_n
     \left({2\pi x/L}\right)\,,
\label{appb:genclass}
\end{equation}
where $h^{(n)}(x)$ is the $n$-th derivative
of $h(x)$ and ${\cal T}_n$ is 
periodic in $x$ with period $L$.
By replacing (\ref{appb:genclass}) in 
(\ref{appb:exaratio}), we find the 
following scaling relation for the dynamical
factor,
\begin{equation}
\frac{\ddot{\gamma}}{\gamma}=
   \frac{1}{L^2}\,{\cal G}_{th}
     \left[\left({W/L}\right)^2,
          {x/L}\right]\,,
\label{fluscal1}
\end{equation}
where ${\cal G}_{th}$ is the theoretical scaling function.

\par
\begin{figure}[htb!]
  \begin{center}
    \begin{tabular}{|c||c|c|} \hline
     \shortstack{\rule{0mm}{1mm}\\
      ${\displaystyle \frac{W}{L}}$\\
         \rule{0mm}{1mm}}
     &\shortstack{\rule{0mm}{1mm}\\$L = 64$\\
         \rule{0mm}{1mm}}
     &\shortstack{\rule{0mm}{1mm}\\$L = 256$\\
         \rule{0mm}{1mm}}					\\ \hline\hline
     \shortstack{${\displaystyle \frac{1}{64}}$\\
       \vspace{1.5cm}}
     &\epsfig{file=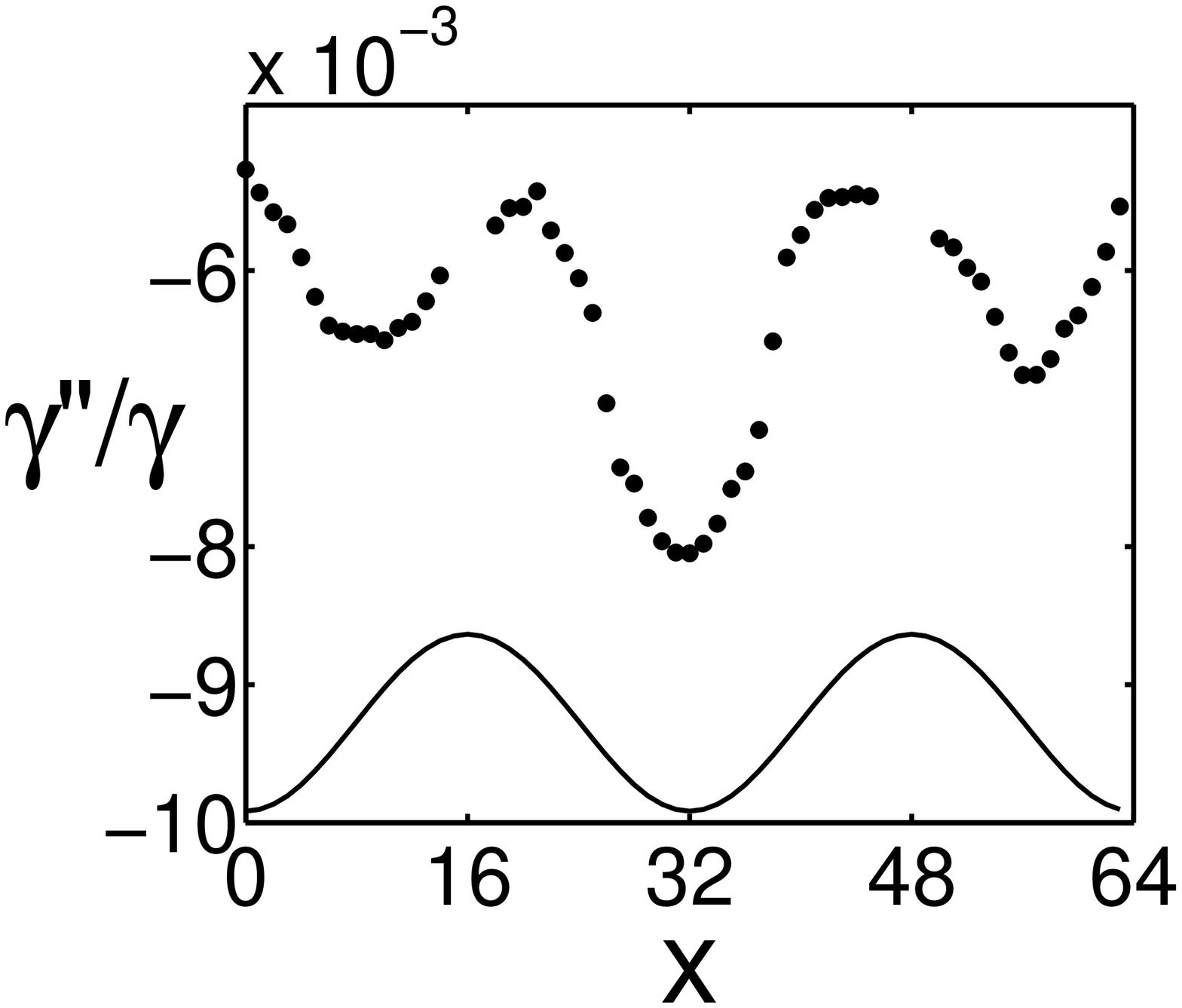,width=.2\textwidth}
     &\epsfig{file=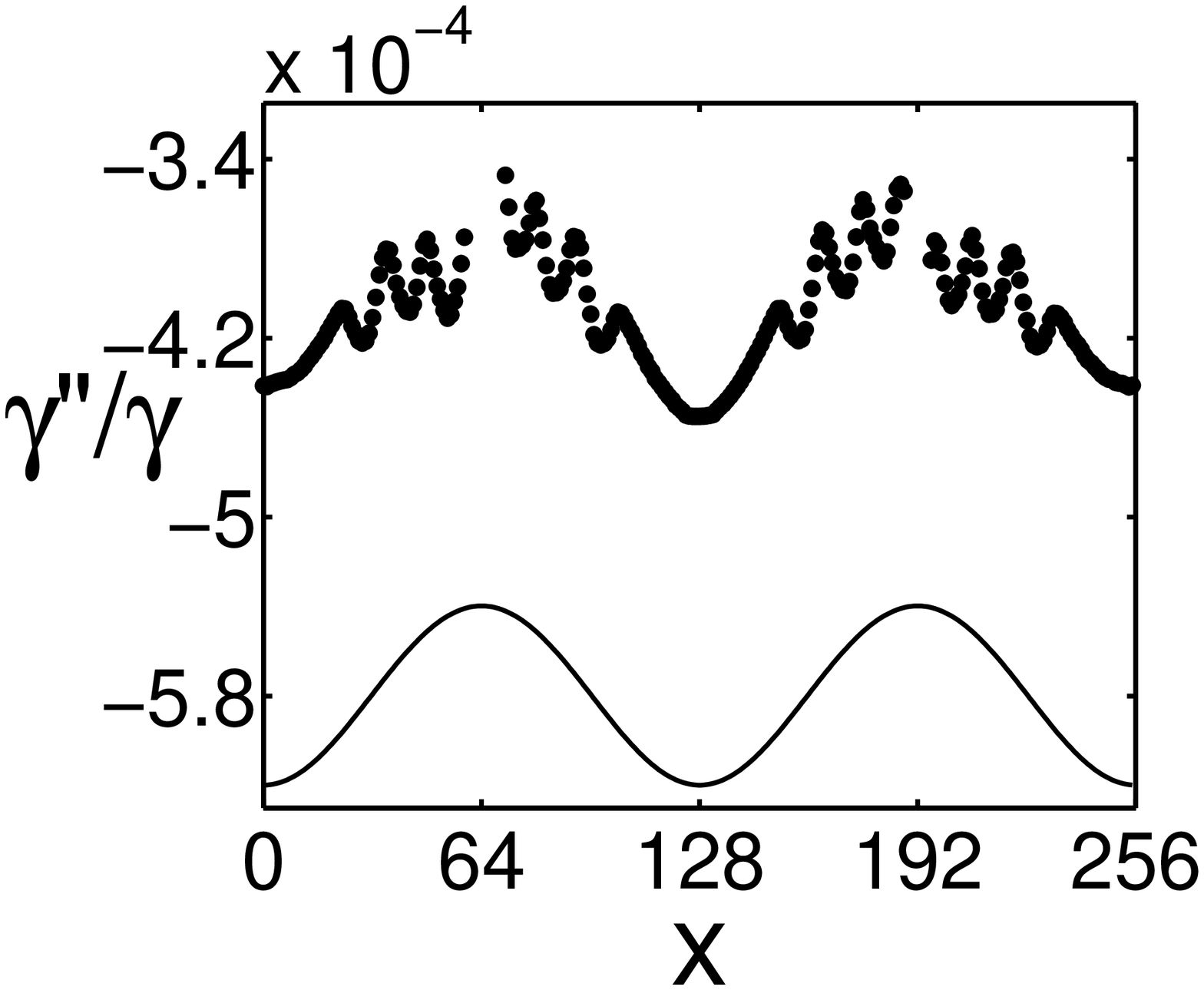,width=.2\textwidth} 	\\ \hline
     \shortstack{${\displaystyle \frac{1}{32}}$\\
       \vspace{1.5cm}}
     &\epsfig{file=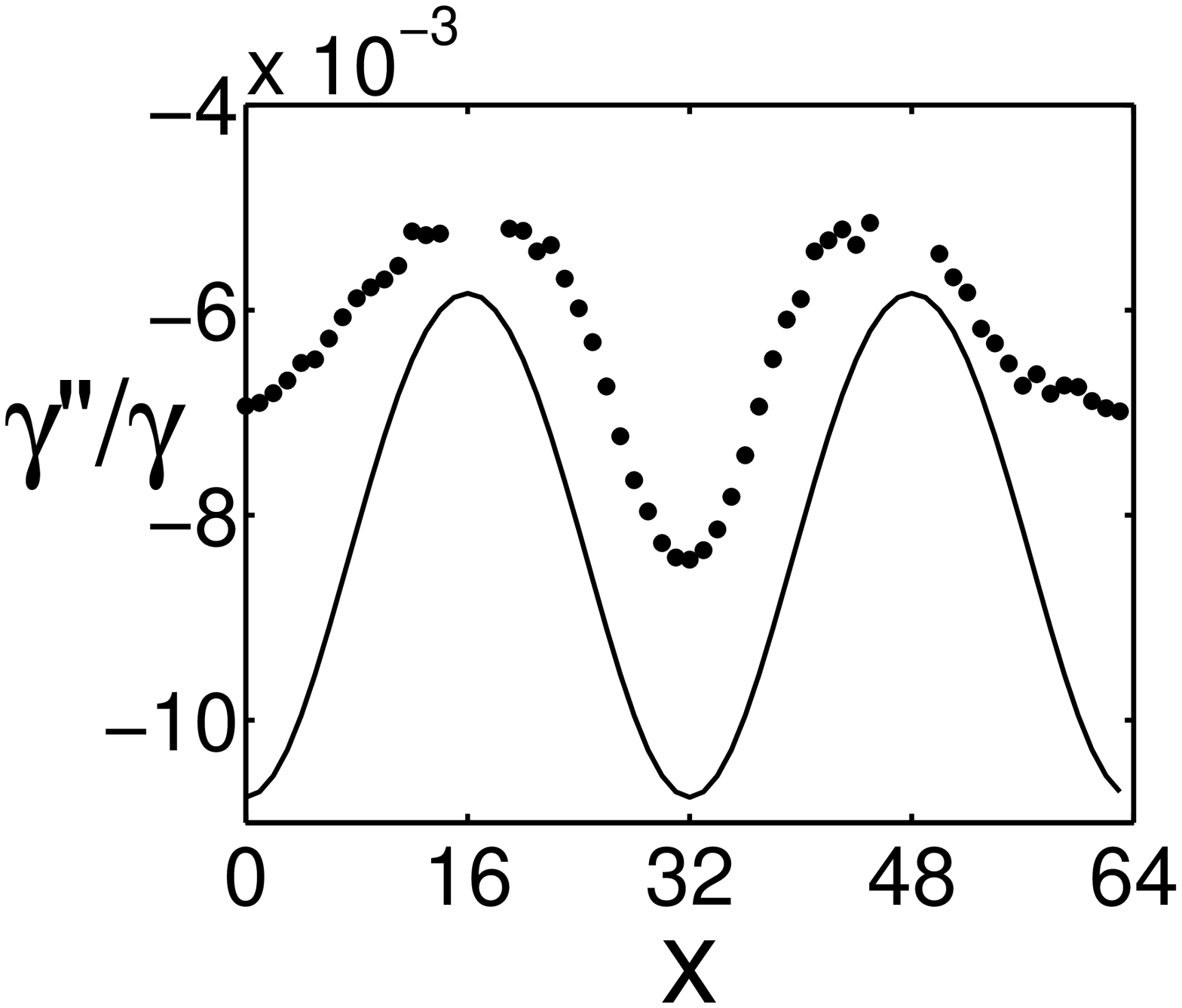,width=.2\textwidth}
     &\epsfig{file=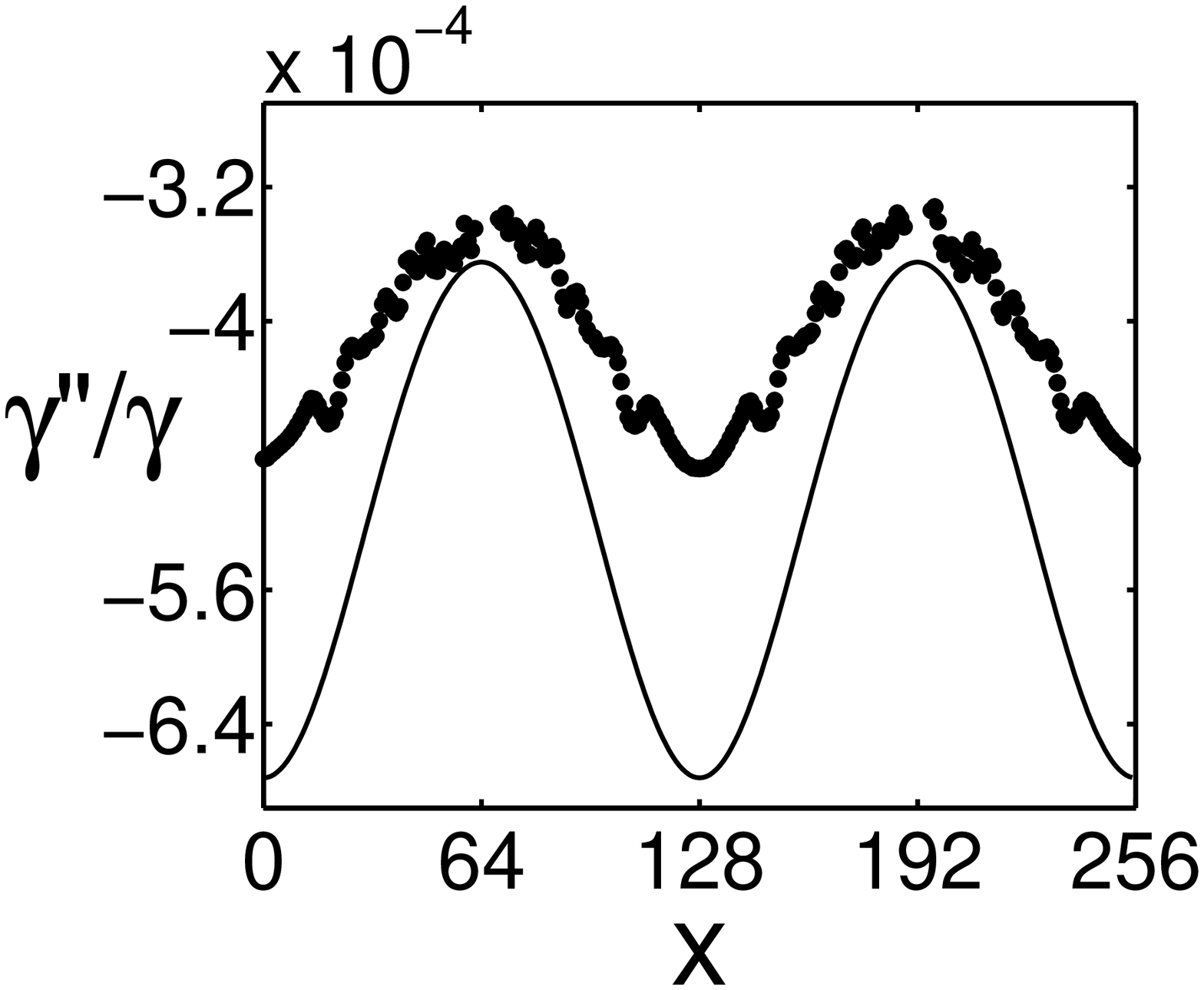,width=.2\textwidth} 	\\ \hline
     \shortstack{${\displaystyle \frac{1}{8}}$\\
       \vspace{1.5cm}}
     & \epsfig{file=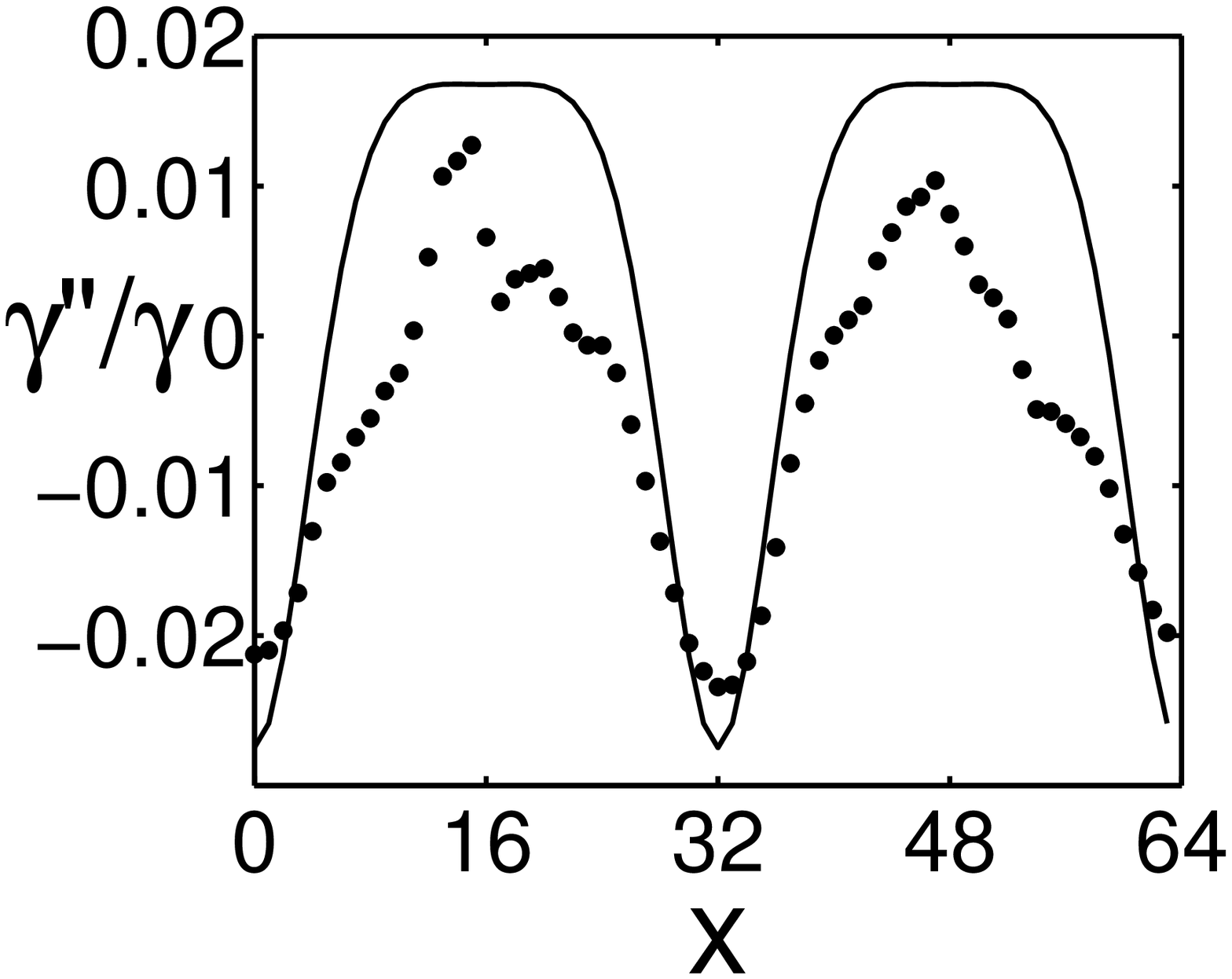,width=.2\textwidth}
     &\epsfig{file=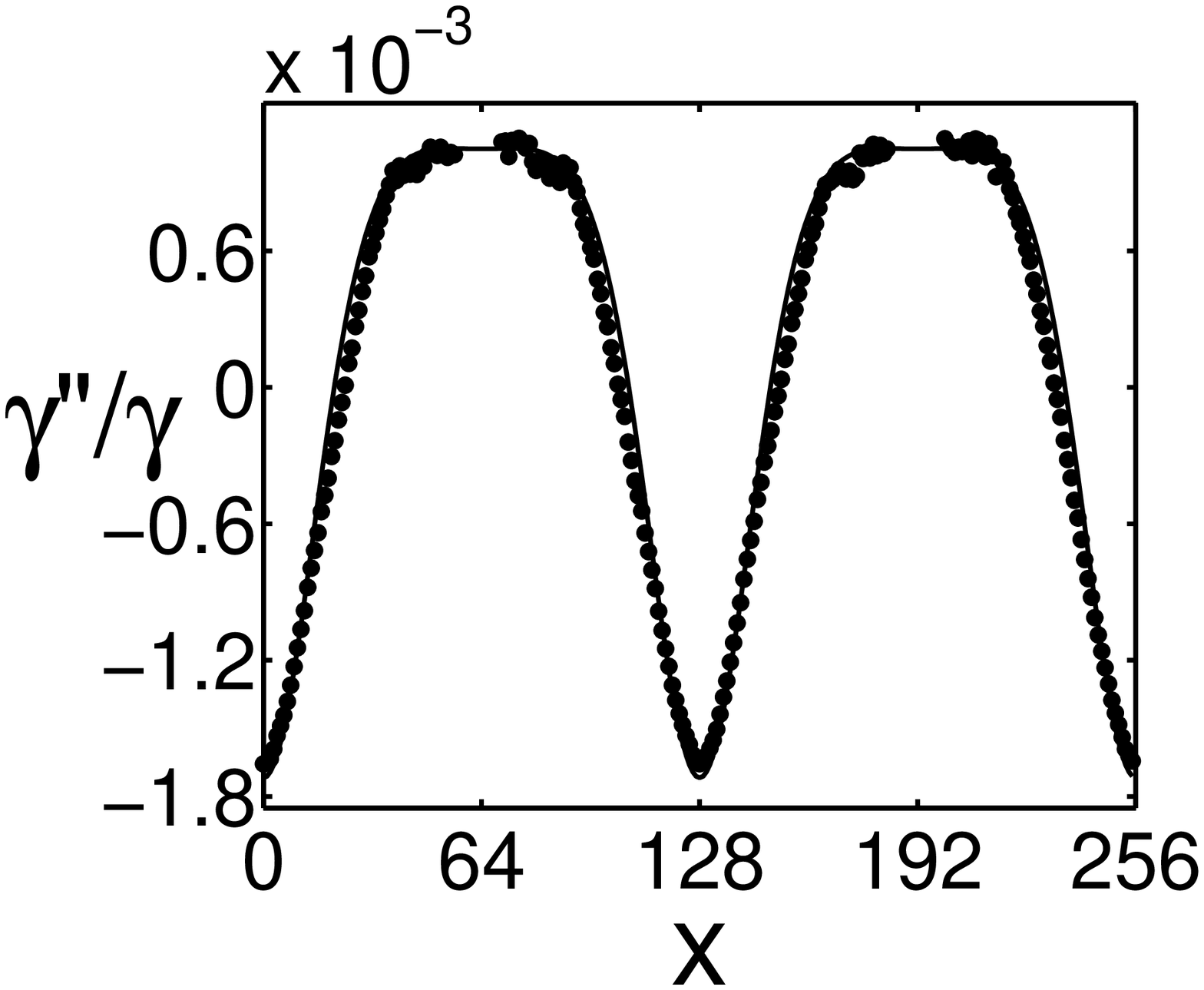,width=.2\textwidth} 	\\ \hline
    \end{tabular}
  \end{center}
  \caption{Measured values of the ratio $\ddot{\gamma}/\gamma$
           are plotted as filled circles for static sinusoidal 
           interfaces with wavelengths $L$ of $64$ 
           and $256$ lattice units and widths $W$ ranging from 1/64
           to 1/8 of the wavelengths. Solid lines represent
           the analytical results for the continuum limit
           obtained from equation (\ref{appb:exaratio}).
           When $W/L < 1/32$, the measured values 
           appear to be shifted with respect to the
           theoretical ones. This introduces an effective 
           surface tension for almost flat surfaces that
           scales as a function ${\cal D}$ of $W$ and $L$: 
           $\sigma_{\rm eff}={\cal D}\!
           \left[\left({W/L}
           \right)^2\right]\mbox{\Large/}L^2$.}
  \label{examp}
\end{figure}

 After initializing the system with different sinusoidal 
waves, we measured the ratio $\ddot{\gamma}/\gamma$
for $L = 64$ and 256 lattice units and
amplitudes ranging from $1/64$ to $1/8$
of the wavelengths.
In Figure \ref{examp} 
we report 
the results of such measurements
as a function of $x$.
From the figure the experimental behavior of 
$\ddot{\gamma}/\gamma$ presents the scaling
\begin{equation}
\left(\frac{\ddot{\gamma}}{\gamma}\right)_{\!\!ex}=
   \frac{1}{L^2}\,
    \left\{{\cal G}_{ex}
     \left[\left({W/L}\right)^2,
          {x/L}\right] + {\cal D}
           \left[\left({W/L}\right)^2
            \right]\right\},
\label{exfluscal}
\end{equation}
where ${\cal G}_{ex}$ approaches ${\cal G}_{th}$ for
large computational boxes (high resolution):
\begin{equation}
\lim_{L\rightarrow \infty}
      {{\cal G}_{ex}\left[\left({W/L}\right)^2,
          {x/L}\right]}=  
{\cal G}_{th}\left[\left({W/L}\right)^2,
          {x/L}\right] \,.
\end{equation}
The most notable feature in (\ref{exfluscal}) is 
the presence of the additional constant term 
${\cal D}[(W/L)^2]/L^2$. It is responsible
for the introduction in our model of an effective surface
tension, as it causes $S$ in (\ref{pert}) to be 
a non-vanishing number.

\subsection{Non-equilibrium roughening}
\label{nonequi}

As shown in Section \ref{introu}, at large times the average 
interface width  reaches its equilibrium value given by 
(\ref{steadywidth}). Before reaching the stationary state, 
though, the interface growth is described by the power-law
(\ref{earlyscaling}). In this section we present the results
of the simulation of non-equilibrium growing interfaces and
compare them with the theoretical predictions.

\begin{figure}[htb!]
  \begin{center}
    \epsfig{file=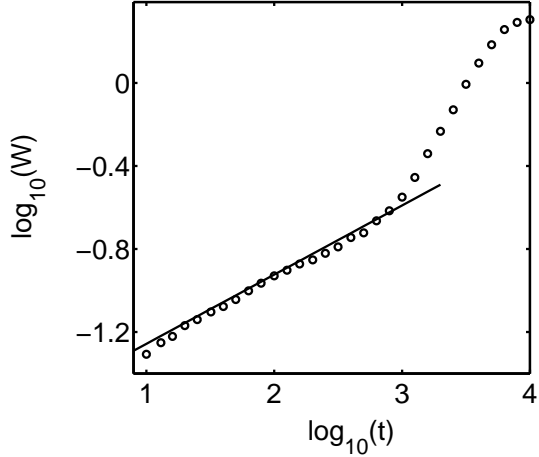,width=.4\textwidth}
    \caption{Log-log plot of the time evolution (circles) of the 
             mean width  of an interface with size $L=64$,
             starting from flatness at time $t=0$, and
             monitored for an interval of time equal to
             about twice the crossover time 
             (\ref{crossovertime}). The solid line 
             represents the theoretical scaling law 
             $W\propto t^{1/3}$ predicted for the 
             effective surface tension 
             $\sigma_{\rm eff}$. The sudden
             change in slope around $t\approx 10^3$
             corresponds to the onset of the purely
             elastic dynamics.}
    \label{fig:flat}
  \end{center}
\end{figure}

\begin{figure}[htb!]
  \begin{center}
    \epsfig{file=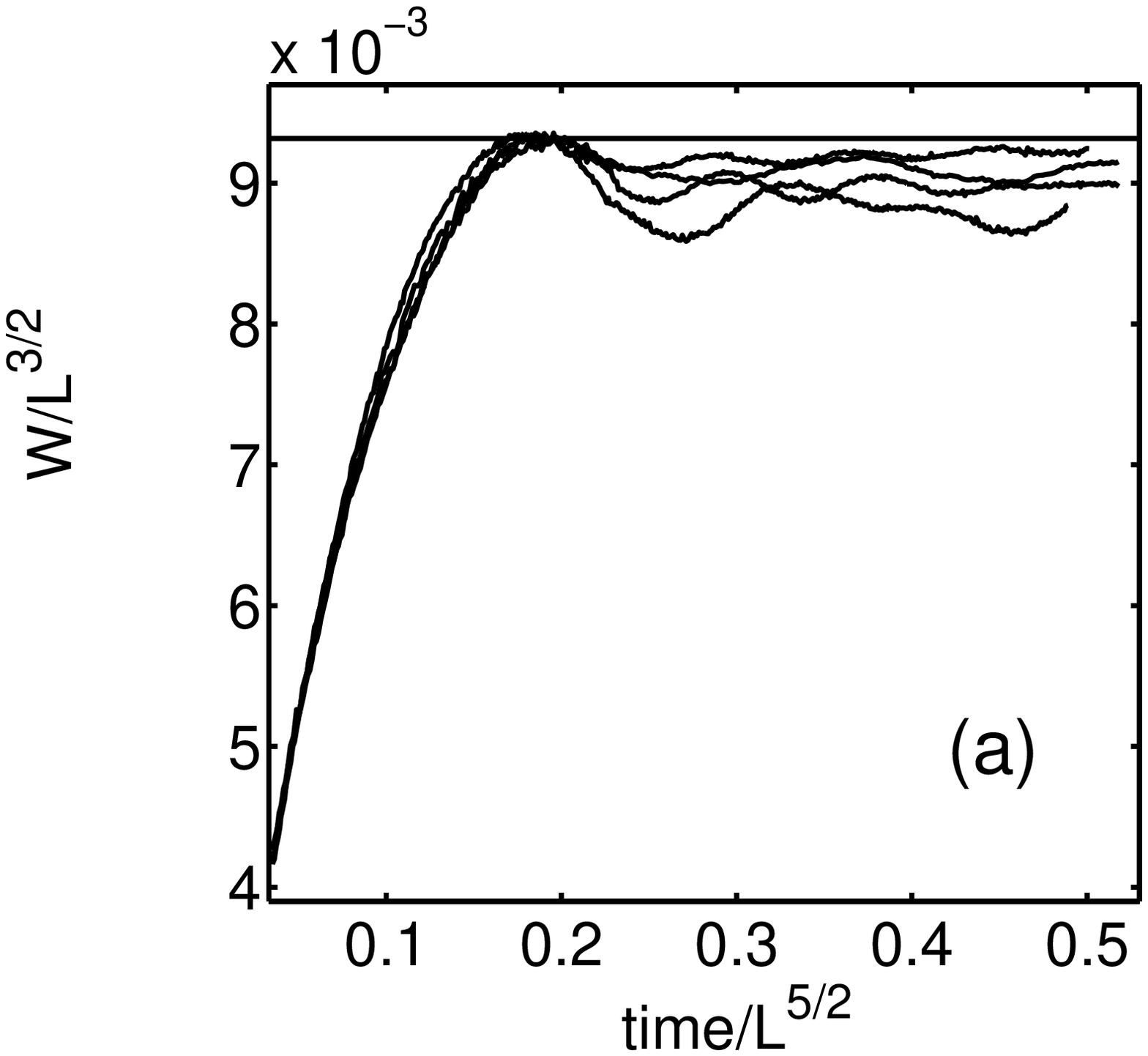,width=.4\textwidth}
    \epsfig{file=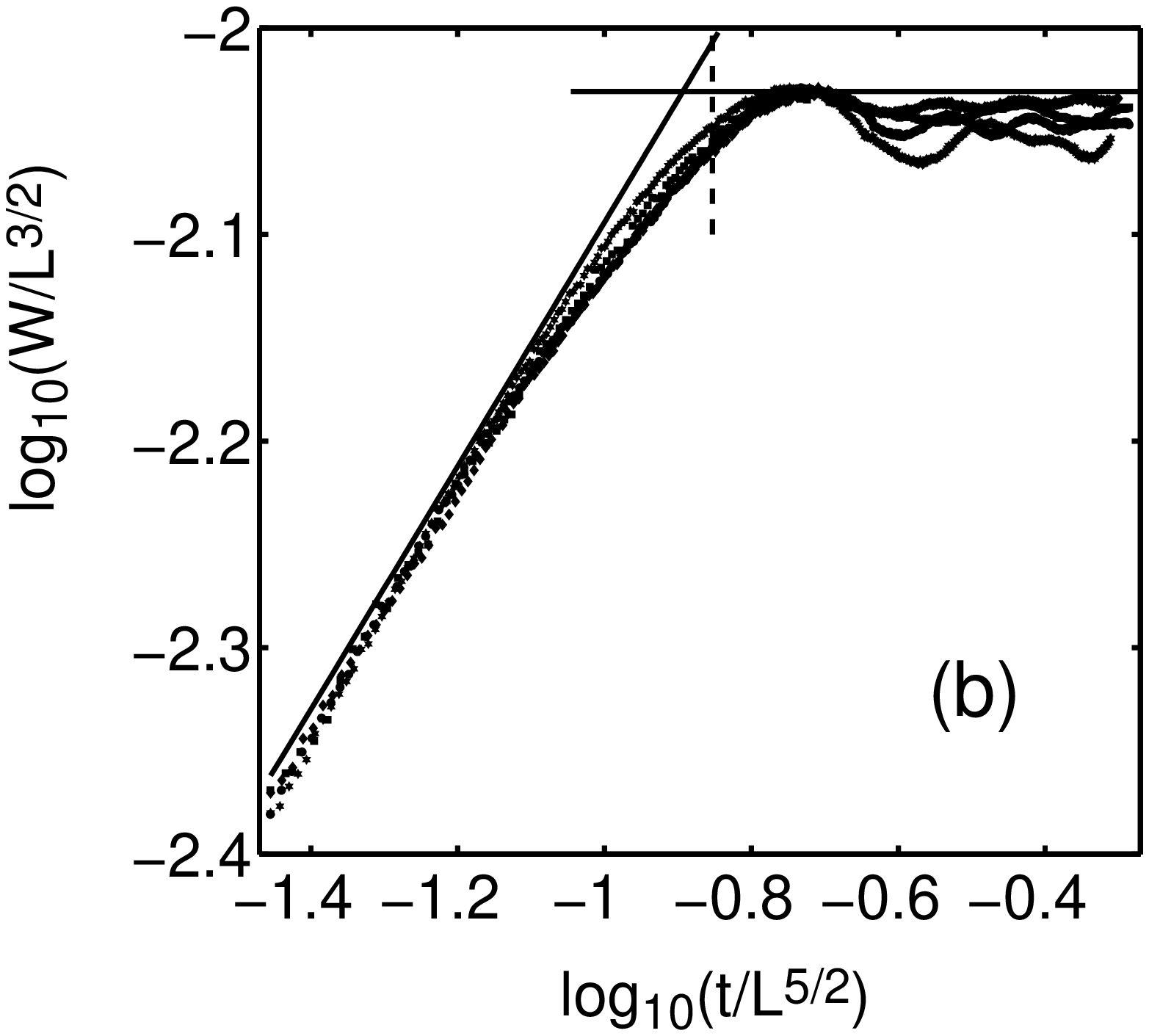,width=.4\textwidth}
    \caption{In (a) we plot the 
             rescaled interface width $W/L^{3/2}$ in lattice units,
             as a function of the time,
             scaled
             to $t/L^{5/2}$ so as to make all the
             crossover times match the prefactor
             $\sqrt{\rho/64\,\pi^3\,\epsilon}$,
             predicted from (\ref{crossovertime}).
             The computational box sizes are 
             $L=32,48,64$ and 96. The horizontal
             line represents the steady state width as 
             in (\ref{steadywidth}).
             In (b) a log-log plot of the 
             same data 
             is given. The dashed line marks the scaled
             crossover time, while the oblique line 
             represents the theoretical prediction from 
             equation (\ref{earlyscaling}), 
             with an angular coefficient of $3/5$.
             The symbols represent different values of the
             system size: squares, diamonds, 
             filled circles and six-pointed stars correspond 
             to $L=32, 48, 64$ and 96 respectively.}
    \label{fig:noneq1}
  \end{center}
\end{figure}

After initializing the system as a flat boundary 
separating two immiscible pre-thermalized fluids, we 
monitored the growth of the interface for a period of 
time approximately equal to twice the crossover time
$t_c$ estimated in (\ref{crossovertime}).
We repeated the simulation  a hundred times and
averaged the results, in order to obtain the
mean quantities that appear in equations
(\ref{steadywidth}) and (\ref{earlyscaling}).
Such an experiment is reported in Figure \ref{fig:flat},
where the solid line represents the expected behavior
of the interface growth, taking into account the 
presence  of the effective 
surface tension.
From the figure, the early-time evolution 
of the interface growth appears 
to be described by the 
scaling law $W\propto t^{1/3}$~\cite{roth4}.
Again, we conclude that the membrane is indeed
driven by $\sigma_{\rm eff}$  for 
very small interface-width/length ratios, that is
for almost flat boundaries.
In order to by-pass the effect of the surface tension
and analyze the purely elastic non-equilibrium
dynamics, the membrane profile is initialized 
in such a way that $W/L>1/20$.
Under these conditions, the disturbance 
due to $\sigma_{\rm eff}$ is negligible
(as shown in Section \ref{section:steady}).
The Fourier components of the interface profile are
prescribed as
$$
h_k(t_o)= \sqrt{\frac{2k_BT}{\epsilon L k^4}}\, 
  \sin [\omega_o(k)t_o]\,,
$$
where $t_o=t_c/4$ is the time at which the 
numerical simulation starts; furthermore,
the relative intensities 
of $h_k(t_o)$ have been chosen so that they
agree with the evolution (\ref{growingphases}).
We studied the non-equilibrium roughening of such
interfaces for an interval of time starting at
$t_c/4$ and terminating at approximately $3t_c$,
after the equilibrium width is attained.
In Figure \ref{fig:noneq1}(a) 
we report the interface
widths for system sizes $L=32, 48, 64$ and 96, 
rescaled as $W/L^{3/2}$ so as to match the prefactor
$\sqrt{k_BT/720\,\epsilon}$ from the anticipated steady 
state (\ref{steadywidth}), drawn in figure as a solid
line.
The horizontal coordinate in Figure \ref{fig:noneq1}(a)
represents the time $t$ in lattice units,
rescaled as $t/L^{5/2}$ so that all the crossover times
fall at $\sqrt{\rho/64\,\pi^3\,\epsilon}$, according
to (\ref{crossovertime}). In order to evaluate
the prefactors above, we used the relations
(\ref{temperature}), (\ref{prop}) and (\ref{visc}),
where the parameters for this experiment were set to
$\lambda_B=-1.5$, $E=0.0002$, $\rho=0.5$ and the 
variance $A$ to $10^{-4}$.
A log-log plot of the same data is showed 
in Figure \ref{fig:noneq1}(b). Also, the predicted
scaling law (\ref{earlyscaling}) is represented
as a solid line with angular coefficient $3/5$.
We notice a good agreement between the numerical 
results and the theoretical predictions, although
we were able to investigate the non-equilibrium
behavior in a time range of just about one-half order
of magnitude.

\section{Conclusions}\label{conc}

Fluctuating elastic interfaces in fluids have been 
discussed from different approaches. 
By means of the fluctuation-dissipation theorem, 
we presented a theoretical derivation of the 
non-equilibrium interface roughening, described
by the scaling law 
$W^2=L^3 \,g\!\left(t/L^{5/2}\right)\,.$
The above expression predicts that the interface,
initially flat, grows as $W\propto t^{3/5}$,
and at long times it reaches the equilibrium width
$W=\sqrt{k_BT/720\,\epsilon}\,L^{3/2}.$
Also, we showed that the correlation function of the 
fluctuations decays exponentially  in the long-time
limit, in agreement with the behavior of the 
correlated forces on the interface, due to the
excitations in the bulk~\cite{roth4}.

Although the theoretical discussion
was necessary for the sake of clarity, this work 
instead focused
on the development of a numerical method simulating 
the phenomenology of fluctuating membranes.
We created a lattice-Boltzmann method.
Starting from the description of the membrane dynamics
by means of the Landau-Helfrich free energy 
\cite{land1,helf}, in which the interface geometry
appears explicitly in the form of the curvature
$\gamma$ and its derivatives, we translated the
macroscopic equation of motion governing the interface
evolution in a perturbation of the single 
relaxation-time lattice-Boltzmann equation.
Such a perturbation depends on $\gamma$, so that a
crucial point of our model is the localization 
of the interface and the measurement of its 
geometric properties. As outlined in Appendix
\ref{appendixb}, we adopted an explicit 
characterization of the boundary followed by a 
polynomial mapping, from which we extracted
the information about the local curvature.

\par We tested our method to reproduce the 
hydrodynamic equilibrium of circular bubbles
and the dynamical coupling of the interface
with the surrounding fluids in the bending wave 
dispersion relation.
Thermal fluctuations were introduced in the model 
by adding a random component to the fluid stress 
tensor~\cite{land2}.
The lattice-Boltzmann equation was thus generalized 
to include a stochastic term~\cite{ladd1}, whose 
fluctuations are uncorrelated in space and time.
The equilibrium frequency power spectrum of 
fluctuating elastic interfaces, predicted by the 
theory and related to the correlation function
of the excitations, was confirmed in our numerical
simulations.
Also, we simulated the non-equilibrium roughening
of membranes, monitoring the time evolution
of the interface width. Its growth rate resulted 
in agreement with the theoretical predictions mentioned
above, although we noticed a disturbance due 
to numerical errors at almost flat surfaces, generating
an unwanted effective surface tension.
In order to improve the effectiveness of the interface
detection process, other, 
more accurate, techniques for tracking the 
interface could be considered. For example, 
using markers in the context of an explicit
discretization of the 
interface~\cite{zal1,zal2} could 
further improve
the performance of this numerical method.

One of the advantages of our numerical method 
relies on the fact that $\gamma$ and 
$d^2\gamma/ds^2$, the second derivative in the
arc length $s$, are not approximated by their
linearized expressions $d^2h/dx^2$ and $d^4h/dx^4$
respectively (here $h(x)$ is the boundary profile).
We could thus use our model to study and simulate
systems in the nonlinear regime, that is when the 
interface is far from being flat and the curvature
is a nonlinear function of the interface profile.
Another feature of this method is that it
has been designed from the outset to describe
thin interfaces separating two (or more) fluids, 
a situation that is difficult to describe 
in terms of a slowly-varying 
continuous field (order parameter).
In future work, we would like to employ this
model in the study of fluctuating membranes
separated in many distinct components, 
colliding with each other and immersed in fluids with 
prescribed or complex fields.
Also interesting would be the inclusion
of a surfactant species,
as in the microemulsion model
of Ref.~\cite{bogho}.

\acknowledgements{We thank Tom Chou and Michael Brenner for valuable
                  discussions. We are also indebted to Tony Dinsmore,
                  Peter Sheridan Dodds and Joshua Weitz 
                  for useful comments.}

\appendix
\section{Fluid-Interface Coupling}\label{appendixa}

In the following we review the dynamical coupling 
between the membrane and the surrounding fluid,
generalizing the results to $n$-dimensional
interfaces. The 2D case with a saddle-splay term 
in the free energy (\ref{freen}) is also
presented.
We shall denote with ${\mathbold r}\equiv (x_1,\dots,x_{n+1})$ the 
locus of the $n$-dimensional hypersurface imbedded in
the $(n+1)$-dimensional space. The hypersurface is 
parameterized by the $n$ coordinates $u_1,\dots,u_n$,
collectively referred to as $u$, so that
${\mathbold r}\equiv {\mathbold r}(u)$.
${\mathbold n}(u)$ represents the unit normal to the surface
oriented outwards and
${\mathbold e}_i \equiv \partial {\mathbold r}/\partial u_i$
is the tangent vector corresponding to the
coordinate $u_i$.
We shall assumes that the hypersurface is well-behaved 
and we are able to choose a parameterization $u$ such
that
\begin{equation}
  \left\langle{\mathbold e}_i,{\mathbold e}_j\right\rangle=
    {\mathbold e}_i\cdot{\mathbold e}_j=\delta_{ij}.
  \label{orto} 
\end{equation}
The Weingarten operator, defined as 
$
L({\mathbold e}_i)\equiv \partial {\mathbold n}/\partial u_i,
$
is self-adjoint with respect to (\ref{orto}), and the 
${\mathbold e}_i$'s are chosen to be eigenvectors of $L$
\begin{equation}
  L({\mathbold e}_i)\equiv 
   \frac{\partial {\mathbold n}}{\partial u_i}=
    \gamma_i {\mathbold e}_i
  \label{eige} 
\end{equation}
where $\gamma_i$ is the principal curvature 
corresponding to ${\mathbold e}_i$.
In other words, our parameterization $u$ runs
along the principal directions of curvature.
We shall apply Hamilton's principle~\cite{gold}
to the line integral of 
the Lagrangian describing the 
interface dynamics
\begin{equation}
 {\cal I}_{id}=
  \int\int{\left( {\rho \over 2}\dot{{\mathbold r}}^2-
    {\epsilon \over 2}H^2 - \sigma\right)dSdt}\,,
  \label{acti} 
\end{equation}
where $\rho$ is the density of the hypersurface,
$\epsilon$ is the bending rigidity, 
$H=\sum_i\gamma_i$ is the mean curvature and 
$\sigma$ the surface tension.
$dS=du_1\dots du_n$ is the volume element for the 
given parameterization.
By means of (\ref{orto}) and (\ref{eige}) one can 
verify that 
\begin{equation}
   \frac{\partial^2 {\mathbold r}}{\partial u_i^2}=
     \frac{\partial {\mathbold e}_i}{\partial u_i}=
       -\gamma_i {\mathbold n},
  \label{nege} 
\end{equation}
so that (\ref{acti}) is recast as
\begin{equation}
 {\cal I}_{id}=
  \int\int{\left[ {\rho \over 2}\dot{{\mathbold r}}^2-
    {\epsilon \over 2}\left(\nabla_u^2 {\mathbold r}\right)^2
     - \sigma\prod\nolimits_{i}{\left|\partial_{u_i}{\mathbold r}\right|} 
       \right]dSdt}
  \label{inr} 
\end{equation}
where
$$
\dot{{\mathbold r}}=\frac{d{\mathbold r}}{dt}, \quad
 \nabla_u^2 {\mathbold r} = 
  \sum_i{\frac{\partial^2 {\mathbold r}}{\partial u_i^2}},\quad
   \left|\partial_{u_i}{\mathbold r}\right|=1\:.
$$
A variation of (\ref{acti}) reads
$$
 \delta{\cal I}_{id}=
  \int\int\left(\rho\dot{{\mathbold r}}
   \cdot \frac{d}{dt}\delta{\mathbold r}-
    \epsilon\nabla_u^2 {\mathbold r}\cdot
     \nabla_u^2 \delta{\mathbold r}\right.
$$
\begin{equation}
 -\sigma\sum_j \frac{\partial{\mathbold r}}{\partial u_i}\cdot
       \left. \frac {\partial}{\partial u_i}\delta{\mathbold r}
       \right)dSdt\,.
  \label{delt} 
\end{equation}
Upon integration by parts, the boundary terms
vanish as we study closed surfaces or 
impose periodic boundary conditions and all of 
the derivatives involved are supposed to be 
smooth functions of the coordinates $u$.
Therefore, by applying Hamilton's principle,
including a term for an external force 
${\mathbold F}_{ext}$
\begin{eqnarray}
 &\delta{\cal I}=
  \int\int{\left[- \rho\ddot{{\mathbold r}}-
    \epsilon \left(\nabla_u^2\right)^2 {\mathbold r}
     +\sigma \nabla_u^2 {\mathbold r} \right]\cdot
      \delta{\mathbold r} dSdt}\nonumber\\
        &+\int{{\mathbold F}_{ext}\cdot\delta{\mathbold r}dt}=0\:,
  \label{leas} 
\end{eqnarray}
one retrieves the following equation of motion
\begin{equation}
  {\mathbold F}_{ext}=-\sigma \nabla_u^2 {\mathbold r}+
    \epsilon \left(\nabla_u^2\right)^2 {\mathbold r}+
    \rho\ddot{{\mathbold r}}  \quad .
  \label{eom} 
\end{equation}
If ${\mathbold F}_{ext}$ is the pressure, it 
is directed along the normal to the surface,
that is ${\mathbold F}_{ext}=F{\mathbold n}$.
By noticing that (\ref{nege}) and (\ref{eige}) lead to
\begin{eqnarray}
 &\left(\nabla_u^2\right)^2 {\mathbold r}=
  (\nabla_u^2) (-H {\mathbold n})=\nonumber\\
  &-\left(\nabla_u^2 H - H \sum\nolimits_i {\gamma_i^2}
   \right){\mathbold n} + \sum_i{\left(2\gamma_i\frac{\partial H}
   {\partial u_i} + H \frac{\partial \gamma_i}{\partial u_i}
   \right){\mathbold e}_i}
  \label{doub} 
\end{eqnarray}
and by projecting $\ddot{{\mathbold r}}$ along the 
normal and tangential components,
(\ref{eom}) can be separated as
\begin{eqnarray}
 & F-\rho\ddot{{\mathbold r}}\cdot{\mathbold n}=
    \sigma H -
    \epsilon \nabla^2 H + \epsilon 
     H \sum\nolimits_i {\gamma_i^2}
     \label{coup}\\
 & \rho\,
     \ddot{{\mathbold r}}\cdot{\mathbold e}_i/\epsilon=
    2\gamma_i\:{\partial H}\!/{\partial u_i} + 
    H \,{\partial \gamma_i}/{\partial u_i}
    \quad i=1\dots n
  \label{components} 
\end{eqnarray}
where we have dropped the subscript $u$ in the 
Laplacian as it is invariant for any isometric
reparameterization.
The main result of this Appendix is that
the dynamical coupling between the interface and the fluid 
in the lattice-Boltzmann method is provided by (\ref{coup}).
Since the macroscopical motion of the membrane is extremely
slow in terms of lattice units, of 
${\cal O}(10^3)$ time steps, the second term in the 
left-hand side of
(\ref{coup}) can be neglected for our purposes and we shall
regard
\begin{equation}
 F=\sigma H -
    \epsilon \nabla^2 H + \epsilon 
     H \sum\nolimits_i {\gamma_i^2} 
  \label{dyna} 
\end{equation}
as a quasi-stationary equilibrium between the
interface configuration and the surrounding
fluid pressure.
Similarly it can be showed that the dynamical coupling
for a 2D interface, including as in (\ref{freen}) 
a saddle-splay correction to (\ref{acti}), is given by
\begin{eqnarray}
  & F=\sigma H -
    \epsilon \nabla^2 H + \epsilon H(H^2 -2K)+\nonumber\\
     & \bar{\epsilon}\left(HK -{\partial^2 \gamma_1}/
       {\partial u_2^2}-
        {\partial^2 \gamma_2}/{\partial u_1^2}
         \right)\,,
     \label{2dcoup}
\end{eqnarray}
where $K=\gamma_1\gamma_2$ is the Gaussian curvature.
(\ref{2dcoup}) should be used in simulating 2D membranes
with this lattice-Boltzmann method, when changes in the 
interface genus are taken into account.

\section{Evaluation of the curvature}
\label{appendixb}

In this appendix we include a description of the 
algorithm used in this work to measure the 
geometric properties of the interface.
The procedure is schematically divided into 
three steps: localization of the interface, 
polynomial approximation, and
evaluation of the local curvature.

As mentioned in Section \ref{sub:micdyn},
the populations corresponding to two immiscible
fluids are distinguished from each other by
splitting the occupation numbers $n_i({\bf x},t)$
at a given lattice site in a ``red'' part,
$r_i({\bf x},t)$, and  a ``blue'' part, 
$b_i({\bf x},t)$, of the distribution function, as in 
(\ref{eq:colors}).
By defining the color field as
\begin{equation}
\label{colorfield}
\Phi({\bf x},t)=\sum_i[r_i({\bf x},t)-b_i({\bf x},t)]\,,
\end{equation}
one can visualize the binary fluid as a 2D surface
where its lowest regions correspond to the physical
presence of the blue fluid and the flat ``highlands''
to the areas occupied by the red fluid. The interface,
in this picture, is represented by the steep slopes.
One can define 
the boundary ${\cal B}$ between the two fluids as the 
set of lattice sites whose
color field absolute value is smaller than a fraction 
$\alpha$ $(<1)$ of the fluid 
density $\rho$, that is
\begin{equation}
\label{boundary}
{\cal B}\equiv \left\{{\bf x}:
      |\Phi({\bf x},t)|<\alpha\rho\right\}\,.
\end{equation}
In our simulations, due to the small thickness of
the interface, $\alpha$ might be chosen to be
a value between 0.5 and 0.9 with no substantial
changes in ${\cal B}$.
Another parameter used in the description of the 
interface is the color gradient ${\bf f}$, defined
as
\begin{equation}
\label{colorgradient}
{\bf f}({\bf x},t)=\sum_i{\bf c}_i\,
    [r_i({\bf x},t)-b_i({\bf x},t)]\,.
\end{equation}
From its definition ${\bf f}$ appears to be, in the lattice
approximation,
perpendicular to the interface.
The vector 
${\bf n}\equiv{\bf f}/|{\bf f}|$ will  be used here  as the unit
normal to the interface.

After localizing the interface, in order to measure
its geometrical properties at a certain
lattice site ${\bf x}_o$ in ${\cal B}$, 
we replace a neighborhood
of ${\bf x}_o$  with
the polynomial resulting from a least-square fit, 
as described below.
The neighborhood of a given boundary 
point (in the present case~${\bf x}_o$) 
is constituted by contiguous lattice sites  in ${\cal B}$ and 
is chosen so that 
the following two
conditions are met:

$\bullet$ The normals at 
the points furthest from~${\bf x}_o$ make an angle of at most $\pi/4$
with the normal at~${\bf x}_o$.

$\bullet$ All the boundary points in the neighborhood lie within
a distance of 30 lattice units from~${\bf x}_o$ (about 10 times the
interface width) or half the linear size of the computational
box, whichever is smaller.

\begin{figure}[htb!]
  \begin{center}
    \epsfig{file=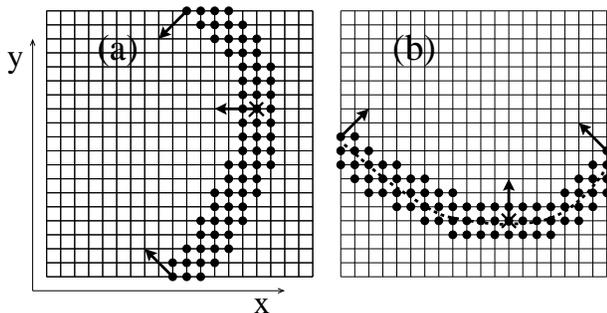,width=.45\textwidth}
    \caption{Filled circles represent an interface neighborhood
             of a given lattice site ${\bf x}_o$ on the boundary
             (marked by a cross). An arrow indicates the
             direction of the normal ${\bf n}$ to the interface at 
             the corresponding lattice site. The whole set of points
             in (a) is rotated clockwise by the angle 
             $\vartheta=\arccos({\bf n}\cdot {\bf \hat x})$ so as to
             align ${\bf n}$ with the $y$-axis (b). 
             A least-square fit of the interface can thus be
             realized by the single-valued polynomial 
             $y=\sum_{i=0}^5a_ix^i/i!$ , here drawn as a dashed line.}
    \label{fig:curva}
  \end{center}
\end{figure}

In Figure \ref{fig:curva} we present the two-step 
process. The filled circles represent the lattice sites 
belonging to the boundary in a 
neighborhood of the point ${\bf x}_o$ (marked
by a cross); arrows are  magnified representations
of the normals at the corresponding lattice sites.
As shown in Figure \ref{fig:curva}(b),
calculations are simplified by rotating clockwise the whole
set of points by the angle $\vartheta=\arccos({\bf n}\cdot{\bf i})$,
where ${\bf i}$ is the unit vector directed along 
the positive $x$-axis. 
In this way, we can fit the subset of ${\cal B}$ by
a single-valued polynomial in $x$: 
$y=\sum_{i=0}^Na_ix^i/i!\;$, where $x=0$ corresponds to the 
horizontal coordinate of ${\bf x}_o$.
It is convenient to choose $N\ge 4$, 
because derivatives up to the fourth
order are involved in the evaluation of
$\ddot \gamma$, the second derivative in the arc length
of the curvature. In practice, we let $N=5$,
as the accuracy of our method does not improve
for larger values of $N$.

Once the coefficients $a_i$ are estimated, we 
calculate $\gamma$ and $\ddot{\gamma}/\gamma$ by 
replacing $h(x)$ respectively in (\ref{curvetta}) and 
(\ref{appb:exaratio})  with 
$y=\sum a_ix^i/i!$ and setting $x$ to vanish.
In other words, $a_n$ substitutes $h^{(n)}$
in (\ref{curvetta}) and (\ref{appb:exaratio}).
The dynamical term 
$\gamma^2({\bf x},t)-\ddot{\gamma}({\bf x},t)/
\gamma({\bf x},t)$, appearing in the 
perturbation (\ref{pert}), is thus evaluated 
by repeating the 
above procedure for each lattice site ${\bf x}$
in ${\cal B}$.

\newcommand{\noopsort}[1]{} \newcommand{\printfirst}[2]{#1}
  \newcommand{\singleletter}[1]{#1} \newcommand{\switchargs}[2]{#2#1}

\end{document}